\documentclass{ws-mpla}

\usepackage{subfig}
\usepackage{xspace}

\newcommand {\rootsNN}  	{\ensuremath{\sqrt{s_{_{NN}}}}}
\newcommand {\roots}    	{\ensuremath{\sqrt{s}}}
\newcommand{\GeVc}			{\ensuremath{{\,\text{Ge\hspace{-.08em}V\hspace{-0.16em}/\hspace{-0.08em}}c}}\xspace}
\newcommand{\MeVc}			{\ensuremath{{\,\text{Me\hspace{-.08em}V\hspace{-0.16em}/\hspace{-0.08em}}c}}\xspace}

\newcommand {\pttrg}        {\ensuremath{p_\mathrm{T}^{\mathrm{trig}}}}
\newcommand {\ptass}        {\ensuremath{p_\mathrm{T}^{\mathrm{assoc}}}}

\begin{document}

\markboth{Wei Li}
{Observation of the ridge in high multiplicity proton-proton collisions}

%%%%%%%%%%%%%%%%%%%%% Publisher's Area please ignore %%%%%%%%%%%%%%
\catchline{}{}{}{}{}
%%%%%%%%%%%%%%%%%%%%%%%%%%%%%%%%%%%%%%%%%%%%%%%%%%%%%%%%%%%%%%%%%%%

%\title{INSTRUCTIONS FOR TYPESETTING MANUSCRIPTS\\
%USING \TeX\ OR \LaTeX\footnote{For the title, try not to use more than 
%three lines. Typeset the title in 10 pt Times Roman, uppercase and 
%boldface.}
%}

\title{Observation of a ``Ridge'' correlation structure in high 
multiplicity proton-proton collisions: A brief review}

\author{\footnotesize WEI LI}
%\footnote{
%Typeset names in 8 pt Times Roman, uppercase. Use the footnote to 
%indicate the present or permanent address of the author.}}

\address{Laboratory of Nuclear Science, Department of Physics, \\
Massachusetts Institute of Technology, 77 Massachusetts Avenue,\\
Cambridge, MA 02139, USA\\
%\footnote{
%State completely without abbreviations, the 
%affiliation and mailing address, including country and e-mail address. 
%Typeset in 8 pt Times Italic.}\\
davidlw@mit.edu}

\maketitle

\pub{Received 8 May 2012}{Revised 30 May 2012}

\begin{abstract}
This paper briefly reviews the striking experimental observation
of a ridge-like dihadron correlation structure in high multiplicity
proton-proton collisions at the Large Hadron Collider (LHC). Recent progress
of both experimental and theoretical efforts on understanding
the physical origin of the novel effect is reviewed. Outlook on
future direction of possible new studies is discussed.
\keywords{ridge; high multiplicity; long-range; near-side}
\end{abstract}

\ccode{PACS numbers: 25.75.Gz, 12.38.Qk, 25.75.Ld}

\section{Introduction}

The observation of a novel long-range dihadron correlation in very high multiplicity
proton-proton (pp) collisions by the Compact Muon Solenoid (CMS) 
collaboration~\cite{CMS_ppridge} at the Large Hadron Collider (LHC) opened up 
the door to a variety of frontiers in the crucial non-perturbative phenomena
of Quantum Chromodynamics (QCD) at a very high density regime. The new finding describes 
a novel correlation that particles coming out of the collision are aligned 
in their azimuthal angle ($\phi$) over a large pseudorapidity ($\eta$) gap ( 
$\eta = -\ln [ \tan(\theta/2)]$ and $\theta$ is the 
polar angle relative to the beam direction). This ``ridge''-like 
structure is found to be absent in minimum bias events but emerges as particle 
multiplicity reaches very high values. This phenomenon has not been observed
before in proton-proton (pp) collisions but resembles similar effects seen in 
collisions of heavier nuclei such as copper and gold ions at the Relativistic 
Heavy Ion Collider (RHIC). Therefore, wide interests have been aroused 
in both the high energy particle and nuclear physics community.

The motivation of studying very high multiplicity hadron production processes is many-sided. 
High multiplicity events are rare in nature and dominated by significant non-perturbative QCD 
activities. Potential new phenomena of QCD could be revealed, and thus warrant 
detailed investigations. Furthermore, with increasing collision rate at the 
highest center of mass energy pp collisions at the LHC, the tail of the multiplicity 
distribution is typically reaching values as high as more than 50 charged particles 
per unit pseudorapidity. Such high particle density begins to approach that of
semi-peripheral collisions of relativistic nuclei such as copper 
at RHIC~\cite{Phobos_bigmult}. Therefore, it is natural to search for the possible signatures of 
high-density and hot QCD matter, known as the ``Quark-Gluon Plasma'' (QGP), in a high multiplicity 
pp environment. The QGP is believed to form in relativistic heavy ion 
collisions with fascinating properties such as close-to-zero shear viscosity over 
entropy density ratio and extremely high opacity~\cite{Arsene:2004fa,Adcox:2004mh,Back:2004je,Adams:2005dq}.

This review is organized as follows: the experimental observation
of the ridge effect in high multiplicity pp collisions at the LHC is first summarized 
in Section~\ref{subsec:ridge_highmultpp}. A brief review of similar phenomena 
in relativistic heavy ion collisions is provided in Section~\ref{subsec:ridge_AA}.
Various theoretical interpretations are described in Section~\ref{sec:theory}.
In the end, a summary and outlook on future directions of possible new studies in order to advance our 
understanding of the ridge is discussed in Section~\ref{sec:summary}.

\section{Observation of the ``Ridge'' in high multiplicity proton-proton interactions}
\label{subsec:ridge_highmultpp}

\begin{figure}[thb]
  \begin{center}
    \includegraphics[width=0.6\textwidth]{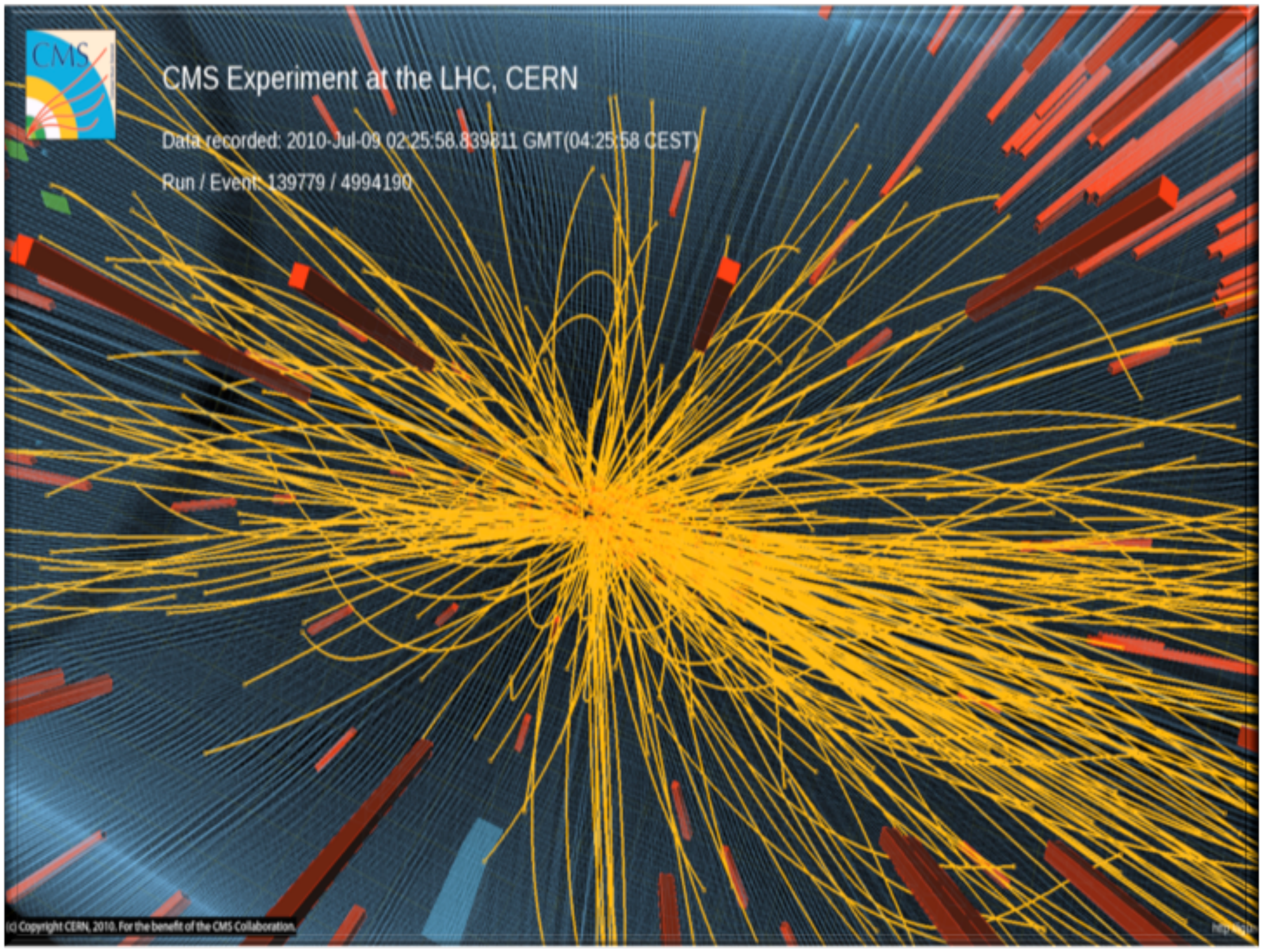}
    \caption{Event display of a single-vertex high multiplicity pp 
    event at \roots\ = 7 TeV recorded by the CMS experiment, where
    there are more than 200 charged particles produced.}
    \label{fig:display}
  \end{center}
\end{figure}

Experimentally, seeking for new phenomena in high multiplicity pp 
collisions is a demanding task. In particular, high-intensity proton 
beams are colliding at the LHC with extremely high rate. The key challenge 
is to identify and record collisions with large number of particles 
originating from a single primary vertex and avoid events from multiple 
low multiplicity pp collisions (pileups). The powerful High-Level Trigger farm in 
CMS ensures the prompt online reconstruction of charged particle trajectories using 
three layers of silicon pixel detector with high resolution in momentum 
(a few \%) and back pointing position ($\sim 100$ microns) such that all primary 
reaction vertices can be precisely located in each bunch crossing.
A display of a very high multiplicity pp event recorded in CMS is illustrated 
in Fig.~\ref{fig:display}, with more than 200 charged particles produced from a single primary 
vertex. A pp data sample corresponding to an integrated luminosity of about 
1 pb$^{-1}$ was used for the first novel observation. 
About 350,000 single-vertex high multiplicity pp events
having average multiplicity 7--8 times that of minimum bias collisions
were recorded. These kind of high multiplicity events are produced with a probability of 
only $10^{-5}$--$10^{-6}$.

\begin{figure}[thb]
    \centering
    \subfloat[]{\includegraphics[width=0.45\linewidth]{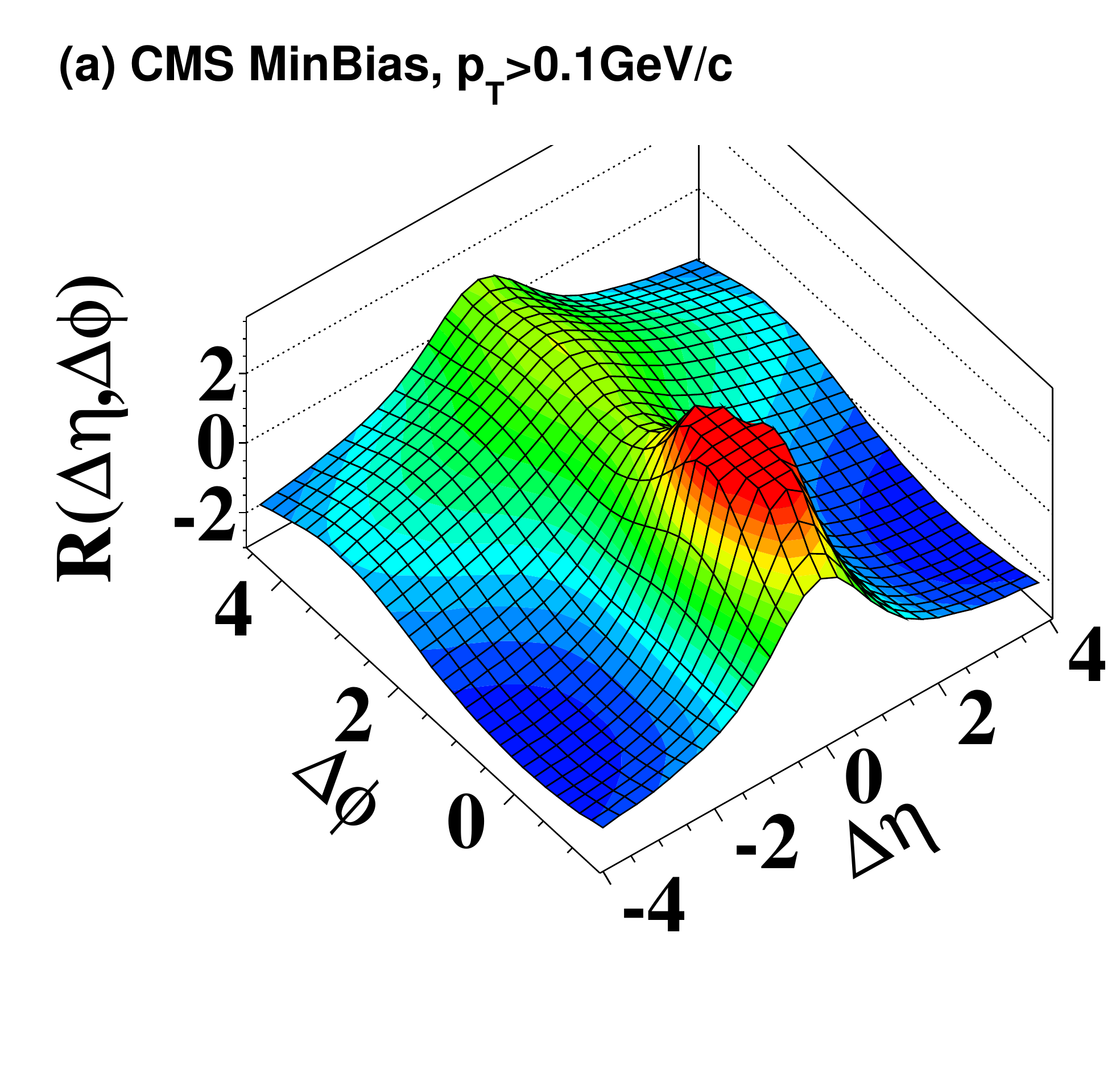}}
    \subfloat[]{\includegraphics[width=0.45\linewidth]{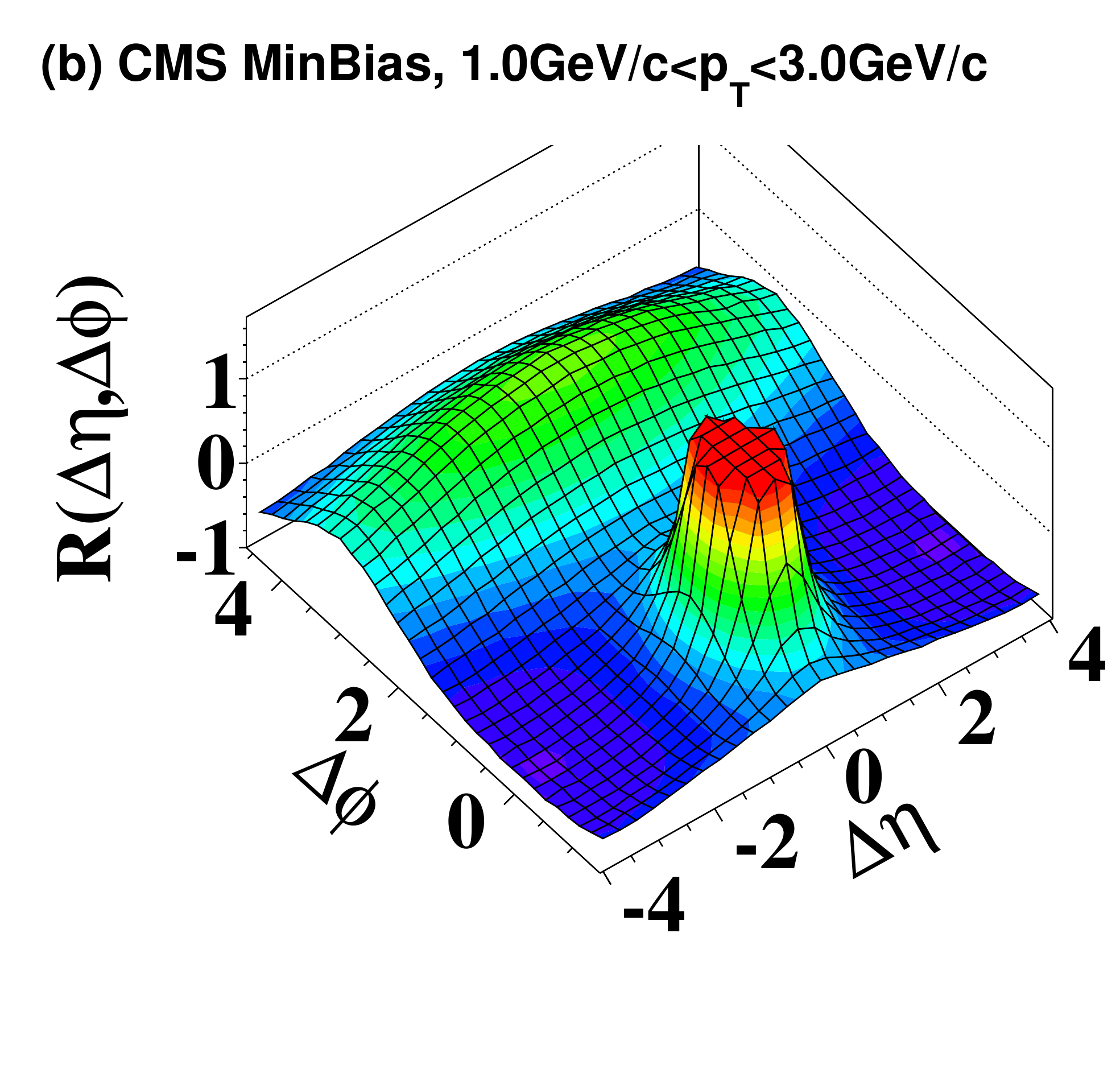}}
    \caption{2-D dihadron correlation functions for minimum bias pp 
    collisions at \roots\ = 7 TeV (a) with $p_T>0.1$\GeVc\ and (b) 
    with $1<p_T<3$\GeVc\ measured by the CMS experiment~\protect\cite{CMS_ppridge}.}
    \label{fig:minbias_2D}
\end{figure}

Studies of multi-particle correlations in pp collisions provide the detailed information on 
the properties of particle production beyond single-particle inclusive
yield measurement. The technique of dihadron angular correlations essentially
reconstructs an imagine of the event structure in the phase space.  
Fig.~\ref{fig:minbias_2D}a shows a transverse momentum ($p_T$) inclusive
dihadron angular correlation function as a function of the relative
pseudorapidity ($|\Delta\eta|=|\eta_1-\eta_2|$) and azimuthal angle 
($|\Delta\phi|=|\phi_1-\phi_2|$) in minimum bias pp collisions at the 
center-of-mass energy (\roots) of 7 TeV for all charged particles with $p_T>0.1$\GeVc, 
measured by the CMS experiment at the LHC~\cite{CMS_ppridge}.
The analysis procedure was established in Refs.~\refcite{Eggert:1974ek,Alver:2007wy,Alver:2008aa}.
The dihadron correlation function is defined as follows:

\vspace{-0.4cm}
\begin{equation}
\label{2pcorr_incl}
R(\Delta\eta,\Delta\phi) =
\left<(\left<N\right>-1)\left(\frac{S_{N}(\Delta\eta,\Delta\phi)}
{B_{N}(\Delta\eta,\Delta\phi)}-1\right)\right>_{bins},
\end{equation}

\noindent where the signal $S_N(\Delta\eta,\Delta\phi)$ denotes
the charged particle pair density and the background 
$B_N(\Delta\eta,\Delta\phi)$ is given by the distribution of 
uncorrelated particle pairs constructed using the event-mixing technique.
Finally, $R(\Delta\eta,\Delta\phi)$ is found by averaging over all event 
multiplicity bins, $N$. The two-dimensional (2-D) 
structure in Fig.~\ref{fig:minbias_2D}a exhibits a variety of features.
A narrow peak at ($\Delta\eta$, $\Delta\phi$) $\sim$ (0,0) is originated 
from higher $p_T$ hard processes like jets; An approximately Gaussian structure 
at $\Delta\eta \sim 0$ (near side) extending over the whole range of $\Delta\phi$
arises from the decay of clusters with lower $p_T$ (e.g., soft QCD string 
fragmentation); In addition, an elongated structure 
(also like a ridge) at $\Delta\phi \sim \pi$ (away side) spreading over a broad 
range in $\Delta\eta$ can be interpreted as due to back-to-back jets or more 
generally momentum conservation. 

\begin{figure}[thb]
  \begin{center}
    \subfloat[]{\includegraphics[width=0.45\linewidth]{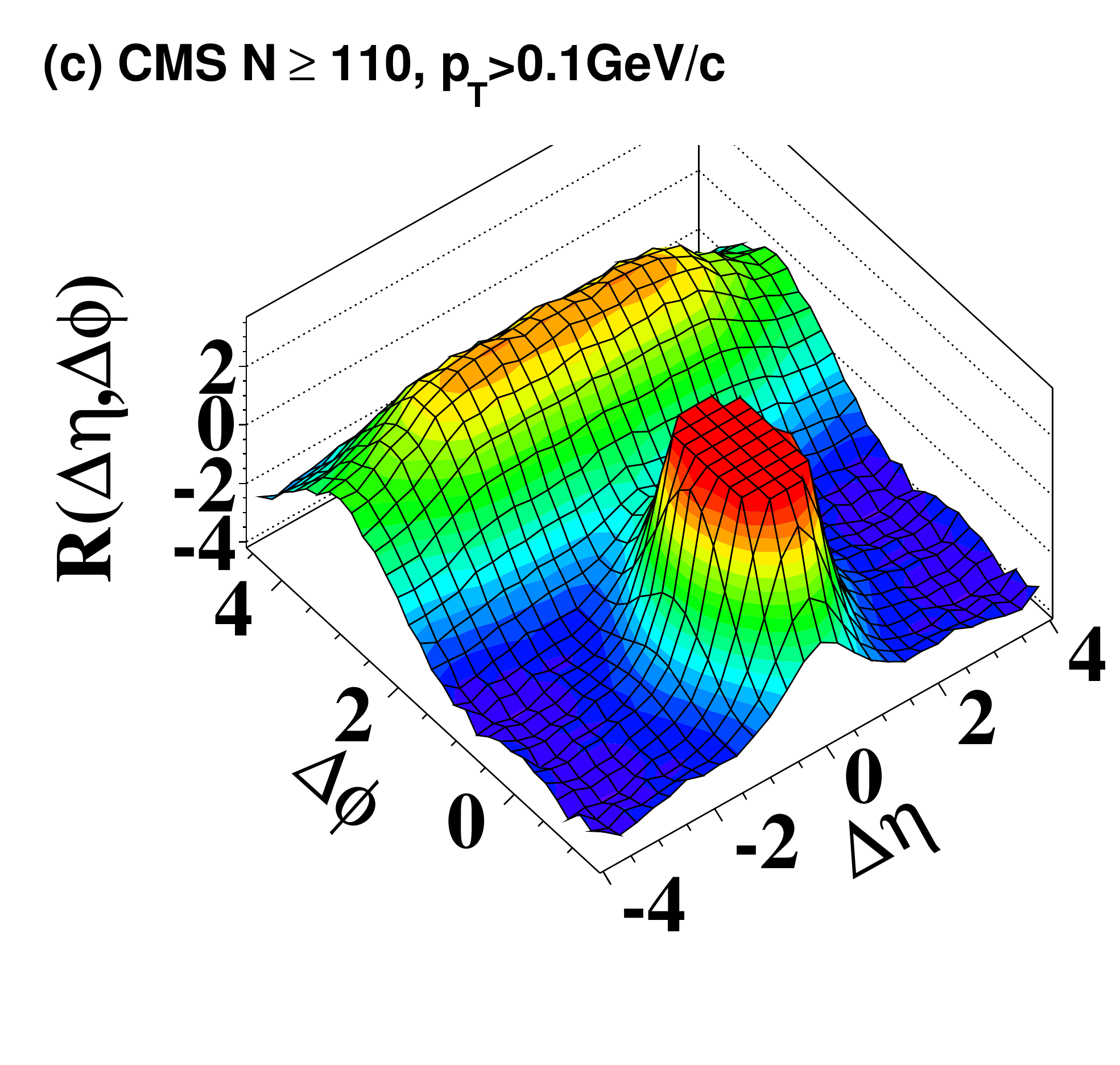}}
    \subfloat[]{\includegraphics[width=0.45\linewidth]{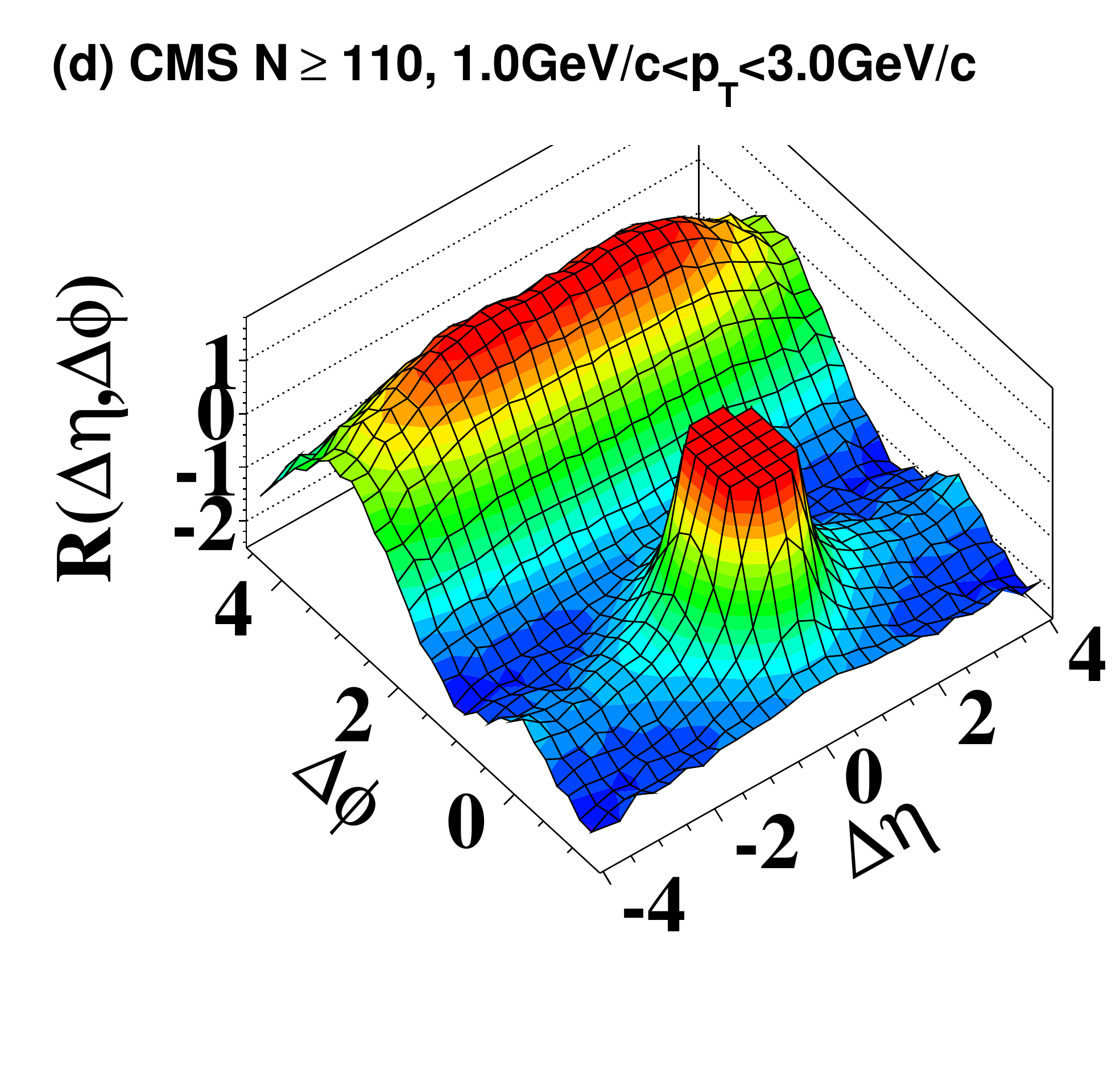}}
    \caption{2-D dihadron correlation functions for high 
    multiplicity ($N \geq 110$) pp 
    collisions at \roots\ = 7 TeV (a) with $p_T>0.1$\GeVc\ and (b) with 
    $1<p_T<3$\GeVc\ measured by the CMS experiment~\protect\cite{CMS_ppridge}. 
    The ``ridge'' refers to the structure in (b) that has a narrow width around
    $\Delta\phi=0$ and extends over the entire $\Delta\eta$ range.}
    \label{fig:highmult_2D}
  \end{center}
\end{figure}

The $p_T$-integrated dihadron correlation function in high multiplicity pp
events, shown in Fig.~\ref{fig:highmult_2D}a, shares a similar structure to that for 
minimum bias events (Fig.~\ref{fig:minbias_2D}a). Besides the more pronounced away-side
correlations due to jettier environment (selection on high multiplicity naturally
biases toward events containing higher $E_T$ jets and enhances the back-to-back correlations), nothing appears to 
be unexpected. Striking phenomena emerges if one not only raises the event multiplicity 
but also varies the transverse momenta of particles. In the intermediate $p_T$ range 
(1--3\GeVc) shown in Fig.~\ref{fig:highmult_2D}b, a striking ``ridge''-like structure 
appears at $\Delta\phi \sim 0$ extending to $|\Delta\eta|$ of at least 4 
units. The ridge is approximately flat in $\Delta\eta$ and also found to have 
no dependence on the charge sign of particles (Fig. 10 in Ref.~\refcite{CMS_ppridge}). 
Same effect can also be observed when correlating an inclusive photon 
(mostly from $\pi^{0}$ decay) with a charged hadron or another inclusive 
photon~\cite{CMS_ppridge}. This novel feature of the data has 
never been seen in dihadron correlation measurements of pp collisions or MC 
generators before.

\begin{figure}[thb]
  \begin{center}
    \subfloat[]{\includegraphics[width=0.45\linewidth]{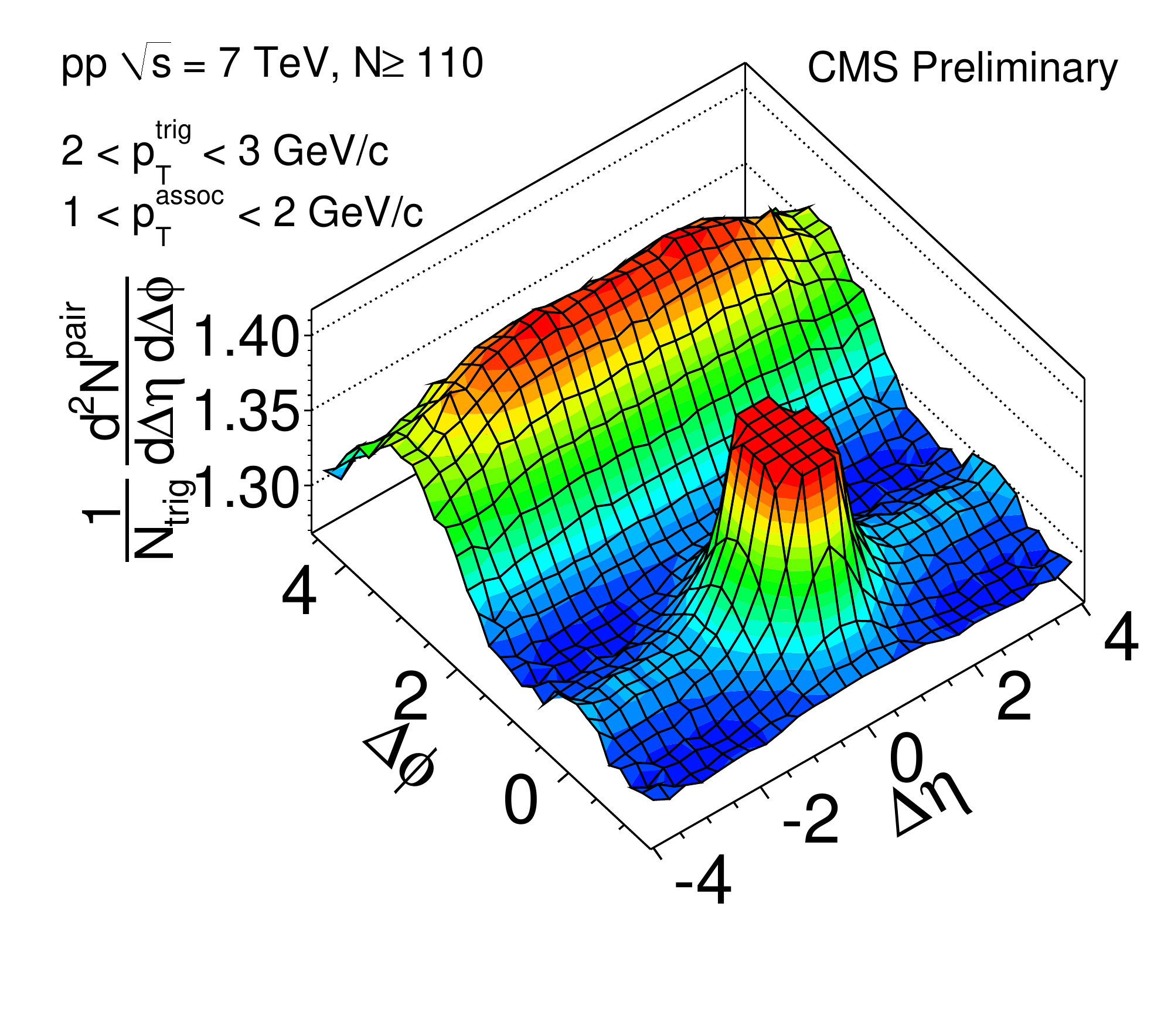}}
    \subfloat[]{\includegraphics[width=0.45\linewidth]{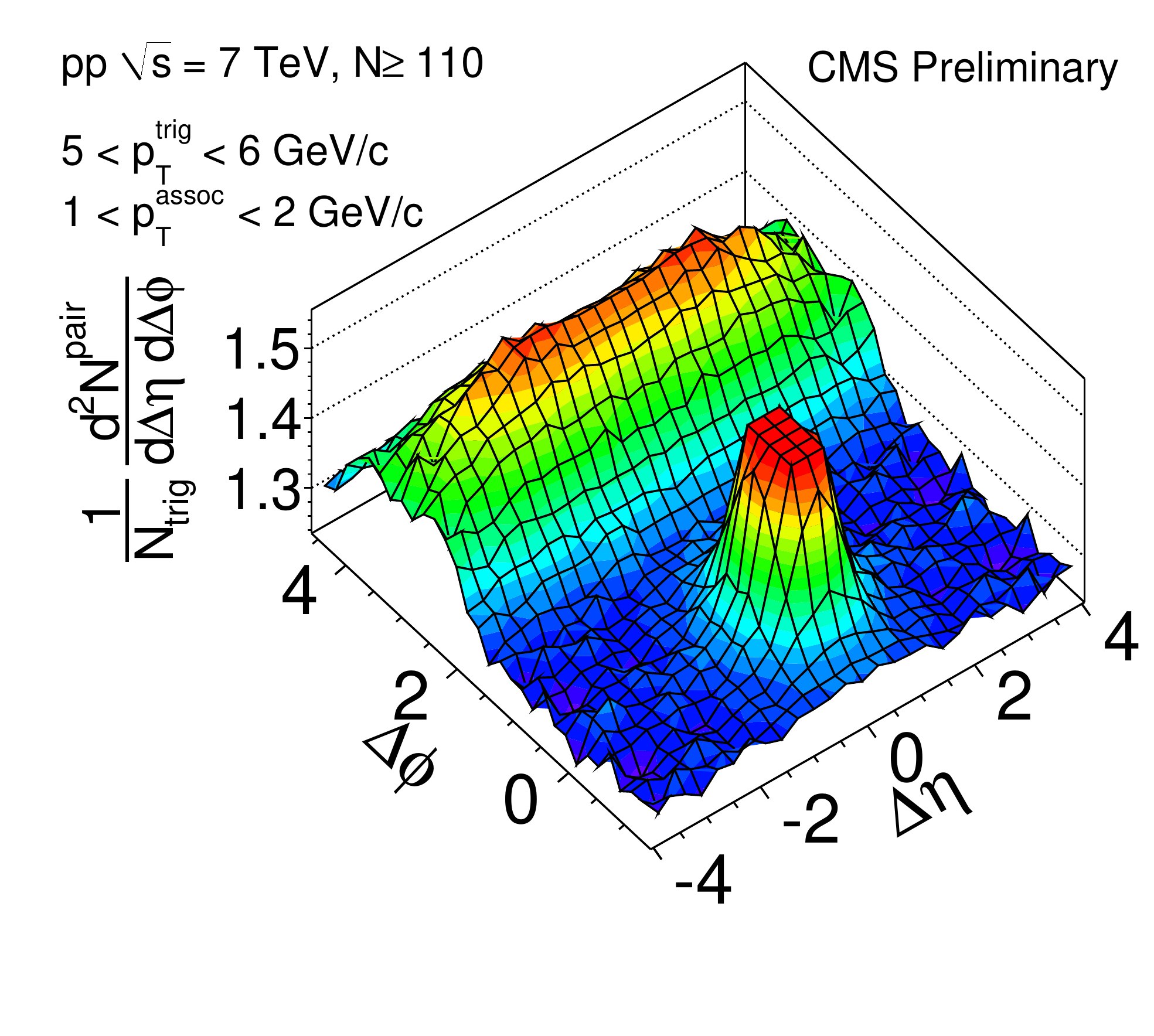}}
    \caption{         
         2-D per-trigger-particle associated yield distribution for charged hadrons
         as a function of $\Delta\eta$ and $\Delta\phi$ from 
         high multiplicity ($N \geq 110$) pp collisions at \roots\ = 7 TeV, for 
         (a) $2<\pttrg<3$\GeVc\ and $1<\ptass<2$\GeVc, and
         (b) $5<\pttrg<6$\GeVc\ and $1<\ptass<2$\GeVc\ measured by 
         the CMS experiment~\protect\cite{HIN-11-006}.}
    \label{fig:highmult_2D_new}
  \end{center}
\end{figure}

Following up the first observation of the ridge in 
high multiplicity pp collisions, the detailed studies of ridge properties
as a function of event multiplicity, transverse momentum and pseudorapidity 
gap were carried out~\cite{HIN-11-006,Li:2011mp}. The analysis was performed
not only for two particles selected from the same $p_T$ range as was done in Ref.~\refcite{CMS_ppridge},
but also for one trigger particle within a specific \pttrg\ range associated
with another particle within a specific \ptass\ range. The \pttrg\ and \ptass\
can be identical or different. The per-trigger-particle associated pair yield distribution,
$\frac{1}{N_{\rm trig}}\frac{d^{2}N^{\rm pair}}{d\Delta\eta d\Delta\phi}$,
as a function of $\Delta\eta$ and $\Delta\phi$ in high 
multiplicity ($N \geq 110$) pp events at \roots\ = 7 TeV for
$2<\pttrg<3~\GeVc$ and $1<\ptass<2~\GeVc$ is shown in 
Fig.~\ref{fig:highmult_2D_new}a. The ridge-like structure is 
clearly visible. However, at higher \pttrg\ of 5--6\GeVc as presented 
in Fig.~\ref{fig:highmult_2D_new}b, the ridge seems to almost disappear.

\begin{figure}[thb]
  \begin{center}
    \subfloat[]{\includegraphics[width=0.45\linewidth]{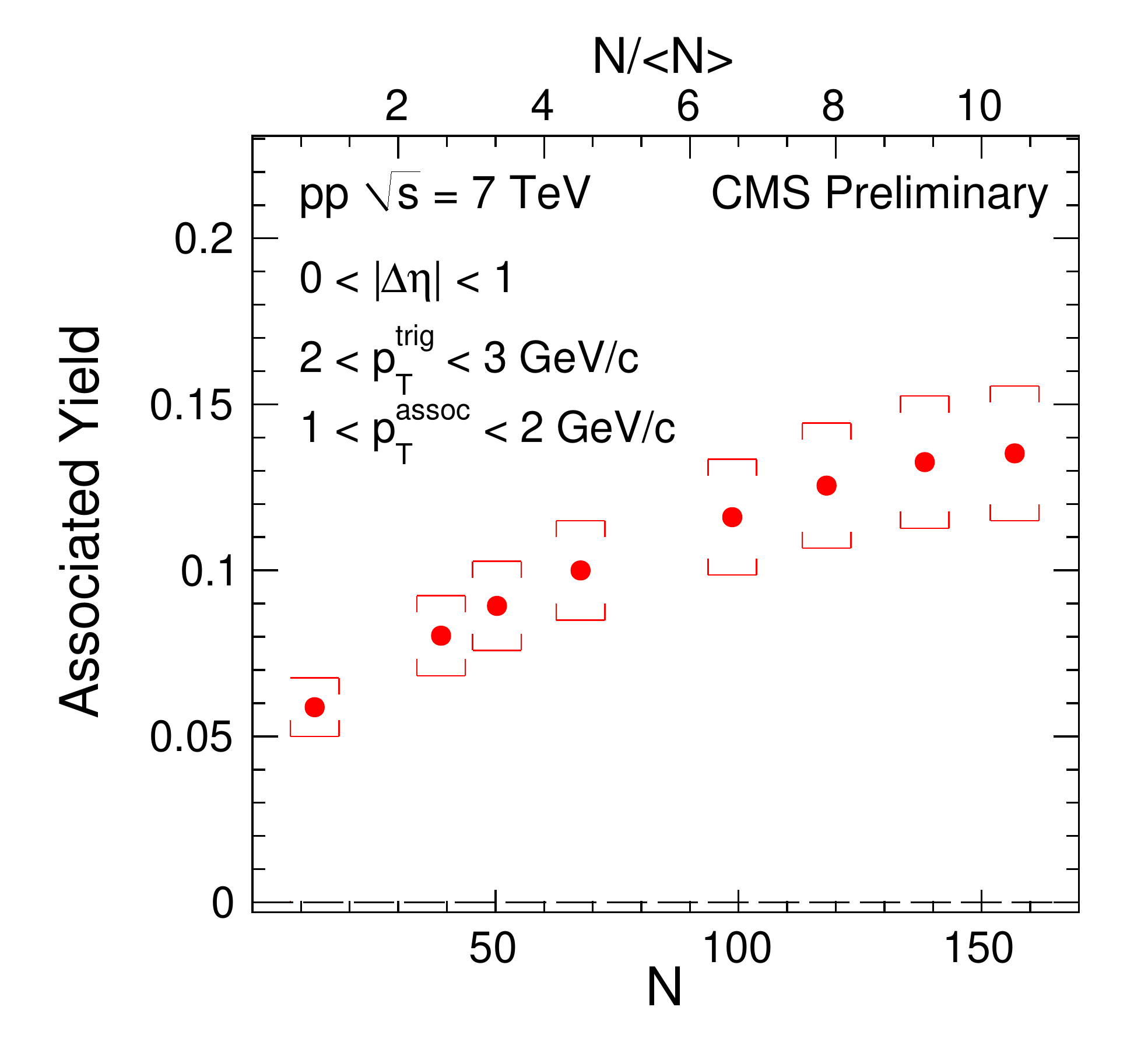}}
    \subfloat[]{\includegraphics[width=0.45\linewidth]{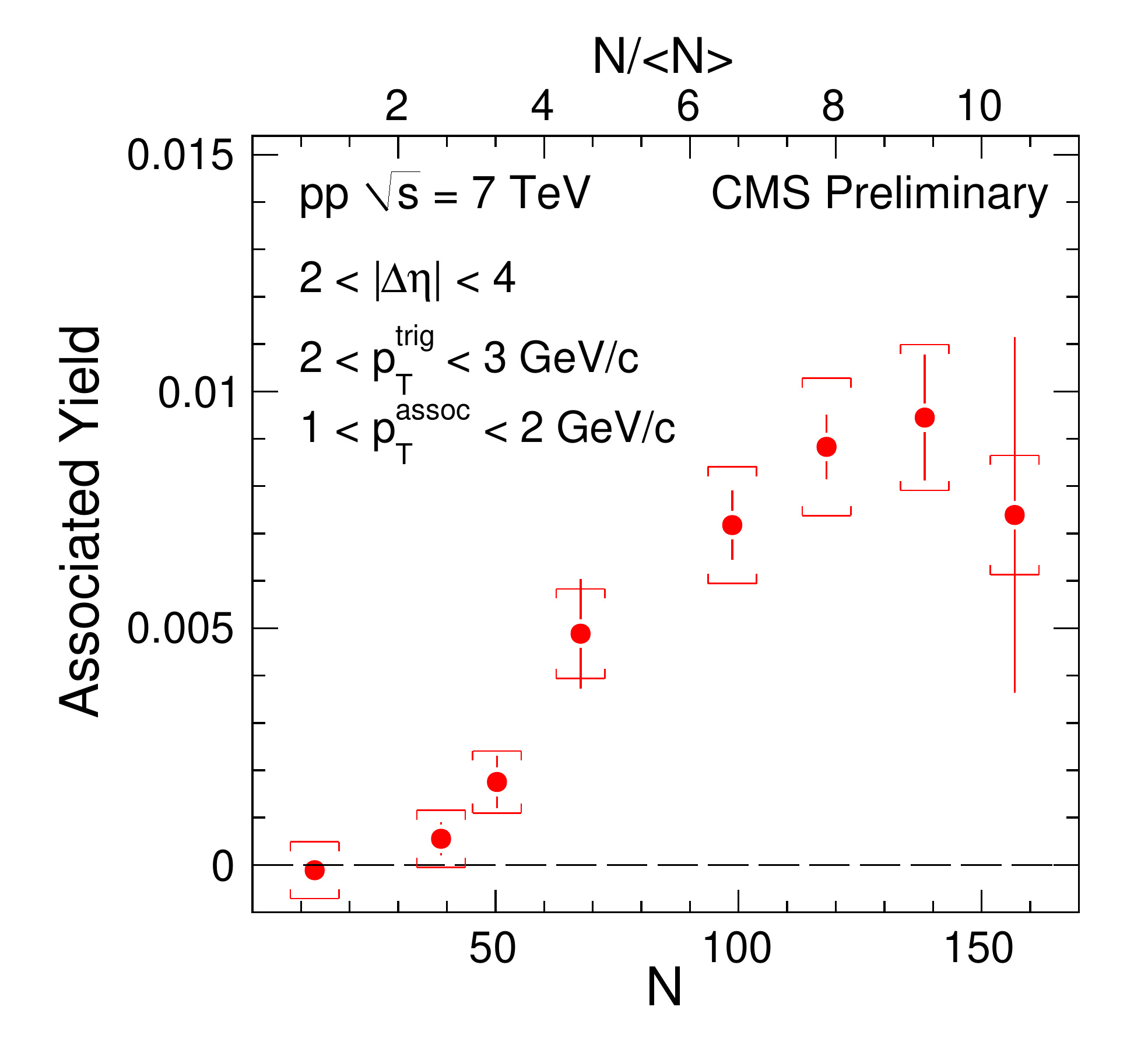}}
    \caption{  
        Integrated associated yields of the near-side ridge in 
        (a) the short-range jet region ($0<|\Delta\eta|<1$) and (b) the 
        long-range ridge region ($2<|\Delta\eta|<4$), for particles with 
        $2<\pttrg<3~\GeVc$ and $1<\ptass<2~\GeVc$, as a function of 
        event multiplicity from pp collisions at \roots\ = 7 TeV measured by 
         the CMS experiment~\protect\cite{HIN-11-006}.}
    \label{fig:yieldvsmult}
  \end{center}
\end{figure}

\begin{figure}[thb]
  \begin{center}
    \includegraphics[width=\linewidth]{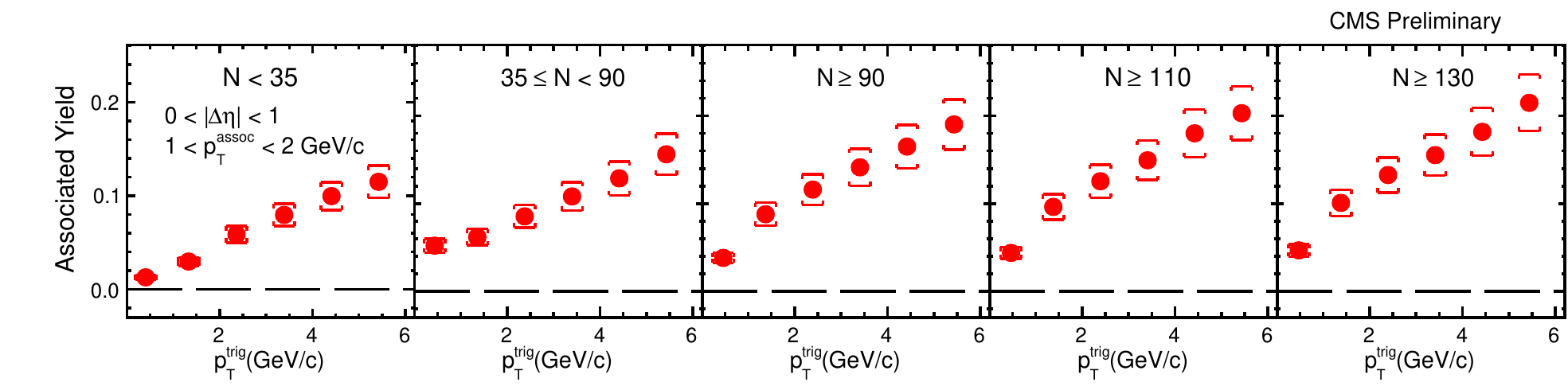}
    \includegraphics[width=\linewidth]{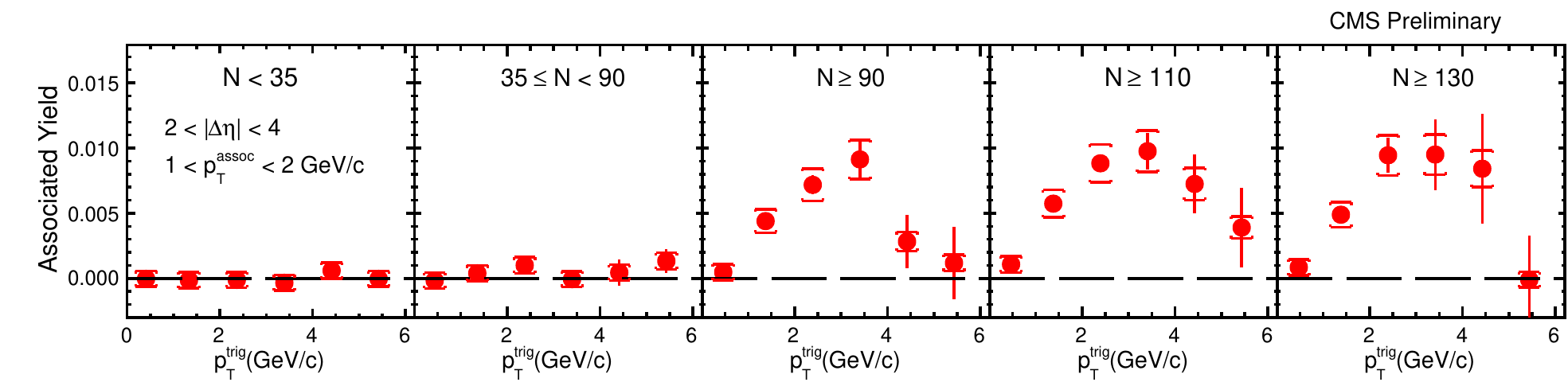}
    \caption{  
        Integrated associated yields of the near-side ridge in 
        (a) the short-range jet region ($0<|\Delta\eta|<1$) and (b) the 
        long-range ridge region ($2<|\Delta\eta|<4$), for particles with 
        $1<\ptass<2~\GeVc$, as a function of \pttrg\ in different bins 
        of event multiplicity from pp collisions at \roots\ = 7 TeV~\protect\cite{HIN-11-006}.}
    \label{fig:yieldvspt}
  \end{center}
\end{figure}

After projecting to one-dimensional (1-D) $\Delta\phi$ correlation functions
in a limited $\Delta\eta$ range, the integrated associated yield is 
calculated on the near side (over the $\Delta\phi$ range from 0 to the minimum $\phi$ position 
of the distribution) for both jet ($|\Delta\eta|<1$) and ridge ($2<|\Delta\eta|<4$) regions
relative to the minimum of the distribution. Fig.~\ref{fig:yieldvsmult}
shows the multiplicity dependence of the near-side associated yield in the 
jet and ridge regions respectively, for the representative transverse momentum bin of $2<\pttrg<3~\GeVc$ and 
$1<\ptass<2~\GeVc$, where the ridge effect appears to be strongest.
The magnitude of the jet yield is enhanced by a factor of roughly 2--3 when going
to events that produce 10 times more multiplicity than minimum bias events.
The ridge effect gradually turns on with event multiplicity around $N \sim 50-60$ (about 
four times of the average multiplicity in minimum bias events) and smoothly increases toward the
high multiplicity region. The \pttrg\ dependence of the jet and ridge yield
is shown in Figure~\ref{fig:yieldvspt} for fixed associated transverse momentum range of $1<\ptass<2~\GeVc$
in bins of event multiplicity. The jet yield increases with \pttrg\ as expected due to 
the increasing contributions from high $E_T$ jets. In high multiplicity events, 
the ridge yield first increases with \pttrg, reaches a maximum around \pttrg\ $\sim$ 
2--3~\GeVc and drops toward even higher \pttrg.

\section{The ``Ridge'' in relativistic heavy-ion collisions}
\label{subsec:ridge_AA}

\begin{figure}[thb]
  \begin{center}
    \subfloat[]{\includegraphics[width=0.32\linewidth]{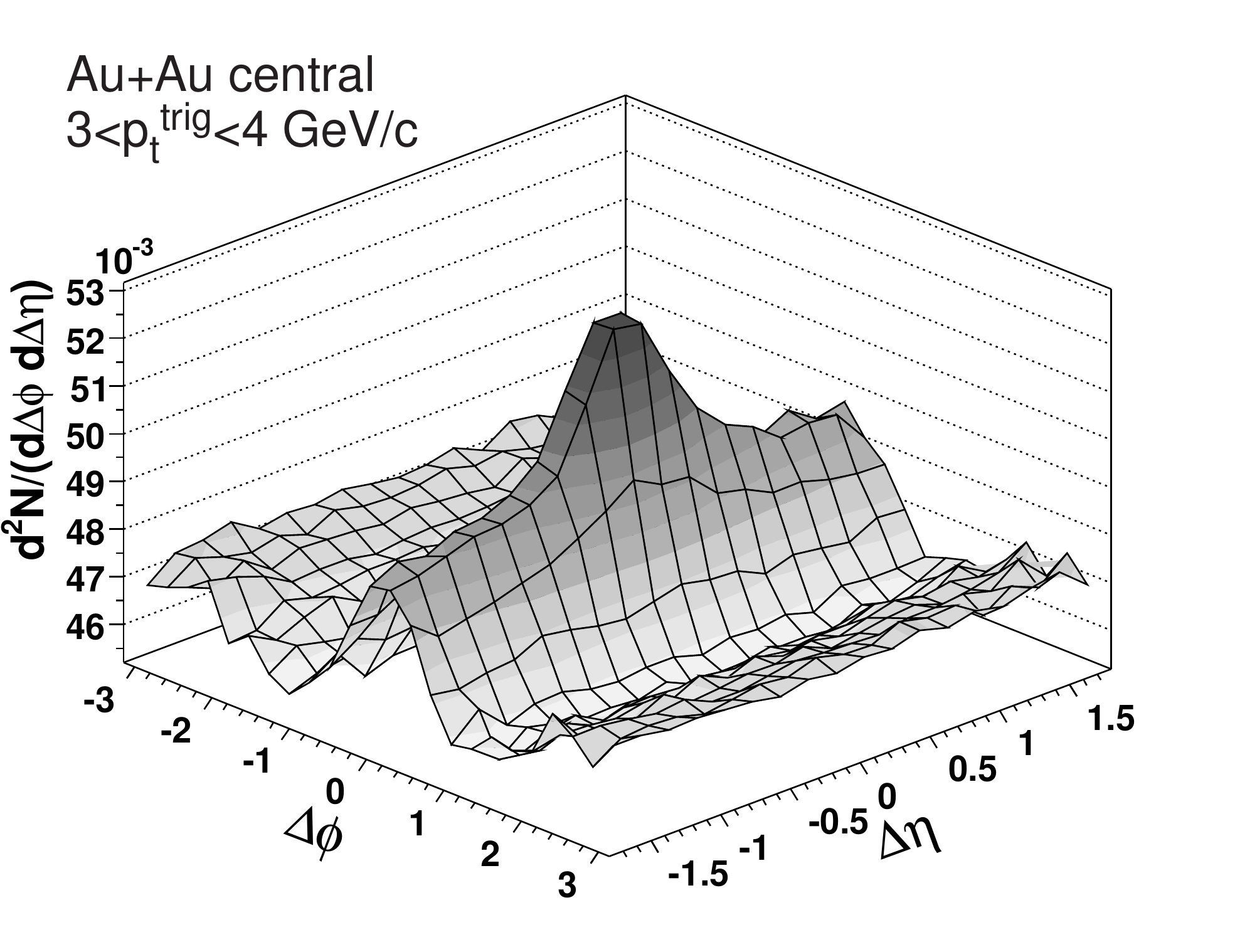}}
    \subfloat[]{\includegraphics[width=0.38\linewidth]{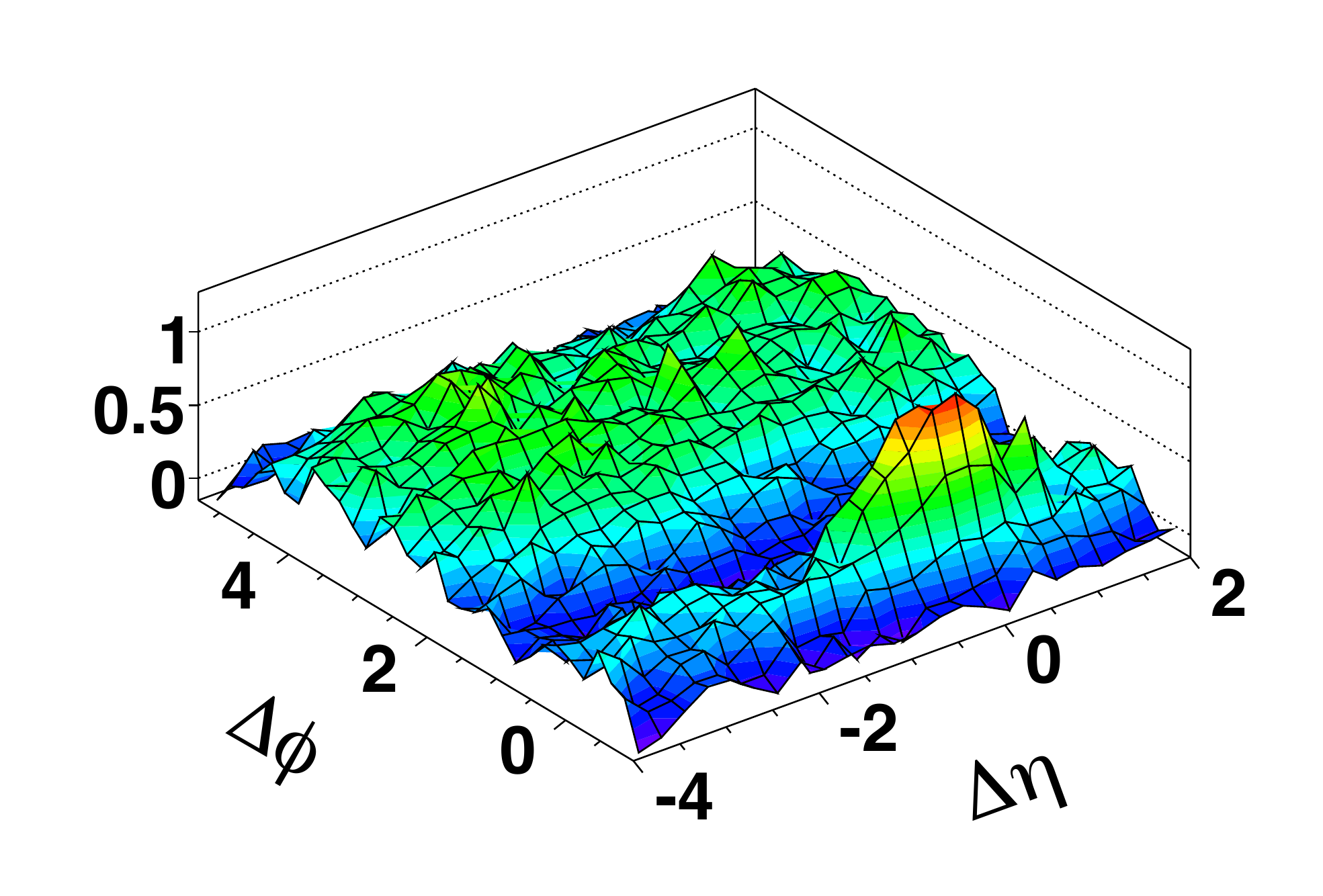}}
    \subfloat[]{\includegraphics[width=0.34\linewidth]{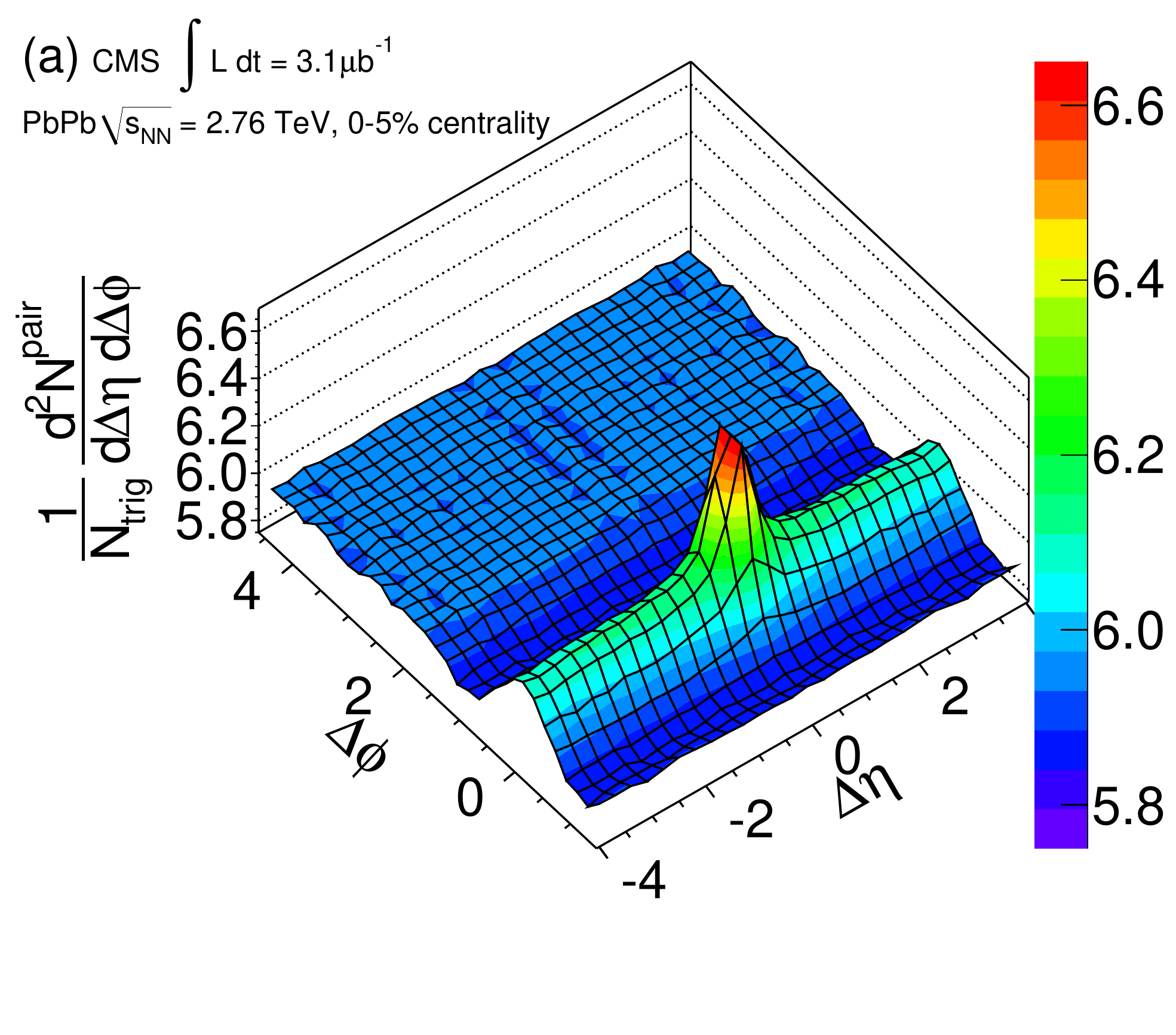}}
    \caption{ 2-D dihadron correlation functions measured in (a) 200 GeV AuAu 
    collisions by the STAR experiment~\protect\cite{Abelev:2009af}, (b) the PHOBOS experiment~\protect\cite{Alver:2009id}
    and (c) 2.76 TeV PbPb collisions by the CMS experiment~\protect\cite{Chatrchyan:2011eka}. 
    The $p_T$ ranges of particles are different for (a), (b) and (c) 
    but not specified here.}
    \label{fig:ridge_HI}
  \end{center}
\end{figure}

Measurement of dihadron correlations is also a powerful tool in 
tackling many aspects of particle production in 
relativistic heavy ion collisions, where properties of particle 
correlations were found to be strongly modified in the presence of a hot 
and dense QGP matter. The near-side ridge structure in 2-D dihadron correlations
was first observed in AuAu collisions at the center-of-mass energy 
per nucleon pair (\rootsNN) of 200 GeV from the STAR experiment 
at RHIC~\cite{Abelev:2009af,Adams:2005ph} (Fig.~\ref{fig:ridge_HI}a).
Note that the ridge studied at RHIC normally refers to the 
residual long-range near-side correlations after subtracting the known
source of correlations from hydrodynamical elliptic flow ($v_{2}$)~\cite{Voloshin:2008dg}.
It was then extended to wider 
$\Delta\eta$ range up to 4 units by the PHOBOS experiment with a wider detector 
acceptance~\cite{Alver:2009id} (Fig.~\ref{fig:ridge_HI}b). The properties of the ridge have been 
extensively studied at both RHIC, and more recently the LHC.
Fig.~\ref{fig:ridge_HI}c shows a representative measurement of dihadron correlations
over a large phase space in 2.76 TeV PbPb collisions from the CMS experiment at the LHC~\cite{Chatrchyan:2011eka},
where a clear and significant ridge-like structure is observed on the near side.
The ridge structure has been first observed for particles 
with transverse momenta from several hundred \MeVc\ to a few \GeVc\
but also found recently to emerge for very high $p_T$ ($>20$\GeVc) particles
at CMS~\cite{CMS_highptridge_PbPb}.

Understanding of the ridge phenomena in heavy ion collisions have 
been evolving over the past several years. It has been qualitatively 
described in many different models~\cite{Armesto:2004pt,Majumder:2006wi,Chiu:2005ad,Wong:2008yh,Voloshin:2003ud,Romatschke:2006bb,Shuryak:2007fu,Dumitru:2008wn,Gavin:2008ev,Dusling:2009ar,Hama:2009vu,Alver:2010gr},
some attributing the ridge to the medium response to its interactions with high-energy partons, while others attribute 
the ridge to the dynamics of medium itself. Motivated by recent theoretical 
developments~\cite{Alver:2010gr,Alver:2010dn,Schenke:2010rr,Petersen:2010cw,Teaney:2010vd},
there has been more and more evidence implying that the $v_2$-subtracted residual long-range ridge effect can 
be understood in analogy to the elliptic flow in the context of hydrodynamics due to the
initial geometric fluctuations, particularly the ``triangularity'', 
leading to higher-order eccentricity and thus final-state azimuthal 
anisotropy~\cite{Alver:2010gr,Alver:2010dn,Schenke:2010rr,Petersen:2010cw,Teaney:2010vd}.
As a result, the long-range dihadron correlations are now commonly analyzed using the technique of 
Fourier harmonic decomposition~\cite{Chatrchyan:2011eka,Chatrchyan:2012wg,Aamodt:2011by,Aad:2012bu}, 

\vspace{-0.2cm}
\begin{equation}
\label{fourier}
\frac{1}{N_{\rm trig}}\frac{dN^{\rm pair}}{d\Delta\phi} \sim \left\{ 1+\sum\limits_{n=1}^{\infty} 2V_{n\Delta} \cos (n\Delta\phi)\right\},
\end{equation}
\vspace{-0.2cm}

\noindent where the Fourier coefficients, $V_{n\Delta}$, are related
to the anisotropy ($v_n$) of final-state particle azimuthal distribution.
The anisotropies for low $p_T$ (up to 1--2 \GeVc) particles are driven 
by the hydrodynamic evolution, while path-length dependence of in-medium parton 
energy loss is believed to be responsible the anisotropy observed for 
higher $p_T$ ($>10$ \GeVc\ or above) particles~\cite{Adare:2010sp,ATLAS:2011ah,CMS_highptv2}. 
Comprehensive studies of dihadron correlations over broad a range of phase 
space and kinematics in heavy ion collisions provide us valuable
information in determining the initial conditions of the QGP matter as well as
medium transport properties such as shear viscosity, opacity etc.
  
\section{Theoretical interpretations of the ``Ridge'' in pp}
\label{sec:theory}

The long-range rapidity correlations in pp collisions were in fact observed
long time ago back to 1980s in the forward-backward multiplicity correlations~\cite{Alpgård1983361}. 
However, what is novel about the ridge in high multiplicity pp is that
particles are produced in a correlated fashion not only over long range in rapidity but also 
collimated in azimuthal angle (near-side). This peculiar new feature was not
observed before in pp or p$\bar{\rm p}$, or any theoretical modelings of pp collisions.

\begin{figure}[thb]
  \begin{center}
    \includegraphics[width=0.7\linewidth]{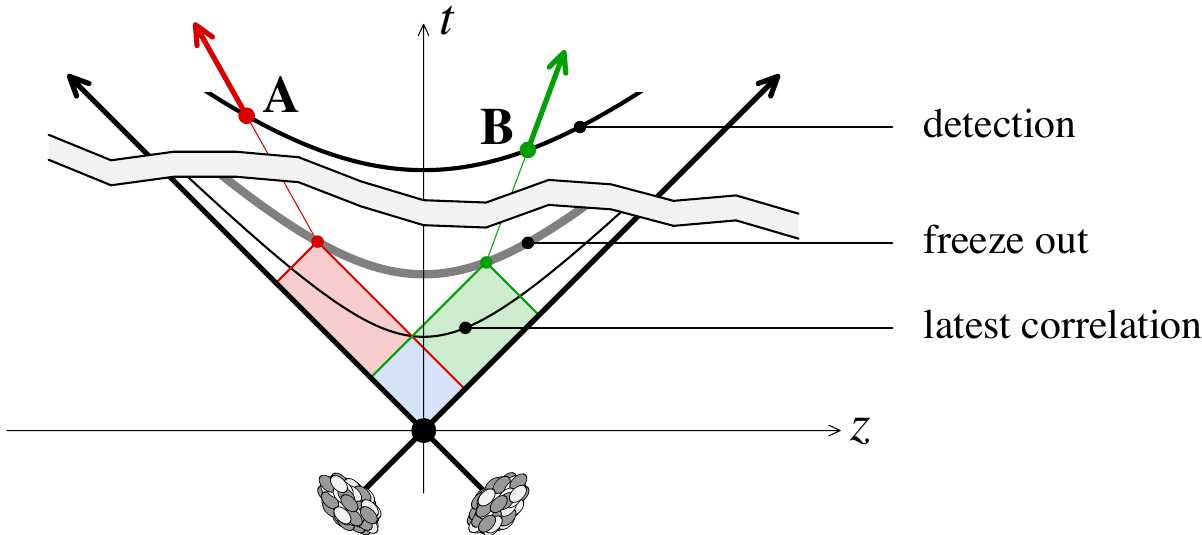}
    \caption{Illustration of causality constraint on the long-range 
    rapidity correlations from Ref.~\protect\refcite{Dumitru:2008wn}, which 
    provides information on early time dynamics.}
    \label{fig:horizon}
  \end{center}
\end{figure}

The microscopic dynamics of high multiplicity particle production and near-side 
ridge correlations in pp have not been fully understood yet.
Nevertheless, such high multiplicity events are likely to result from 
very ``central'' pp events (small impact parameter, or large overlapping region like 
the central nucleus-nucleus collisions), where multiparton 
interactions become more relevant~\cite{CasalderreySolana:2009uk,Strikman:2011cx}. 
Furthermore, based on causality constraints of particle production as illustrated 
in Fig.~\ref{fig:horizon} from Ref.~\refcite{Dumitru:2008wn},
any final-state correlation over large rapidity gaps (several units) has to be
established shortly after the interaction happens at proper time earlier than

\begin{equation}
\tau_\textrm{init.} = \tau_{\rm freeze-out}\, \exp\left(-\frac{1}{2} |\Delta y| \right) \,,
\end{equation}

\noindent which is strongly suppressed at large rapidity gap, $|\Delta y|$~\cite{Dumitru:2008wn}.
Here $\tau_{\rm freeze-out}$ is the freeze-out time of particle production.
An acute analogy of it is the large scale correlations and fluctuations
in the cosmic microwave background (CMB). The fact that the structure of the 
universe today is so smooth (up to one part in 100,000) and correlated
over large distances that are casually disconnected indicates the existence of
a rapidly inflationary era in the very early stage of the universe. 
The amount of correlations and fluctuations observed in the CMB
provide crucial information on the primordial quantum fluctuations, which
lead to the formation of galaxies. Therefore, the observation of long-range
near-side ridge correlations in pp interactions gives us an exciting opportunity
to investigate the initial-state structure of the proton wavefunction, as well
as the quantum fluctuations of the color field at very short timescale.

As already pointed out in Sec.~\ref{subsec:ridge_AA}, the near-side 
ridge phenomena in heavy ion collisions can be interpreted in terms of
the hydrodynamic phenomena. The eccentricity of initial-state energy 
density profile in the transverse plane of the overlapping nucleus-nucleus 
system is propagated to the final-state particle azimuthal anisotropy 
via pressure driven radial flow on an event-by-event basis. Since these 
anisotropies are originated from the initial condition, they 
are boost invariant in rapidity. In dihadron angular
correlations, the hydrodynamic flow effect emerges in the form of Fourier harmonic 
components, $\sim cos(n\Delta\phi)$, referred to as the elliptic ($n=2$) and 
higher-order ($n>2$) flow. Each of the Fourier terms gives rise to a local maximum at $\Delta\phi=0$ 
independent of $\Delta\eta$, which sums up to a ridge-like structure. In principle, a 
similar ridge structure should also be present on the away side ($\Delta\phi \sim \pi$) 
of the correlation function. However, it is often mixed up with the so-called
``non-flow'' correlations from dijets that are back-to-back 
in $\phi$ extending over $\Delta\eta$, thus is more complicated.

\begin{figure}[thb]
  \begin{center}
    \subfloat[]{\includegraphics[width=0.45\linewidth]{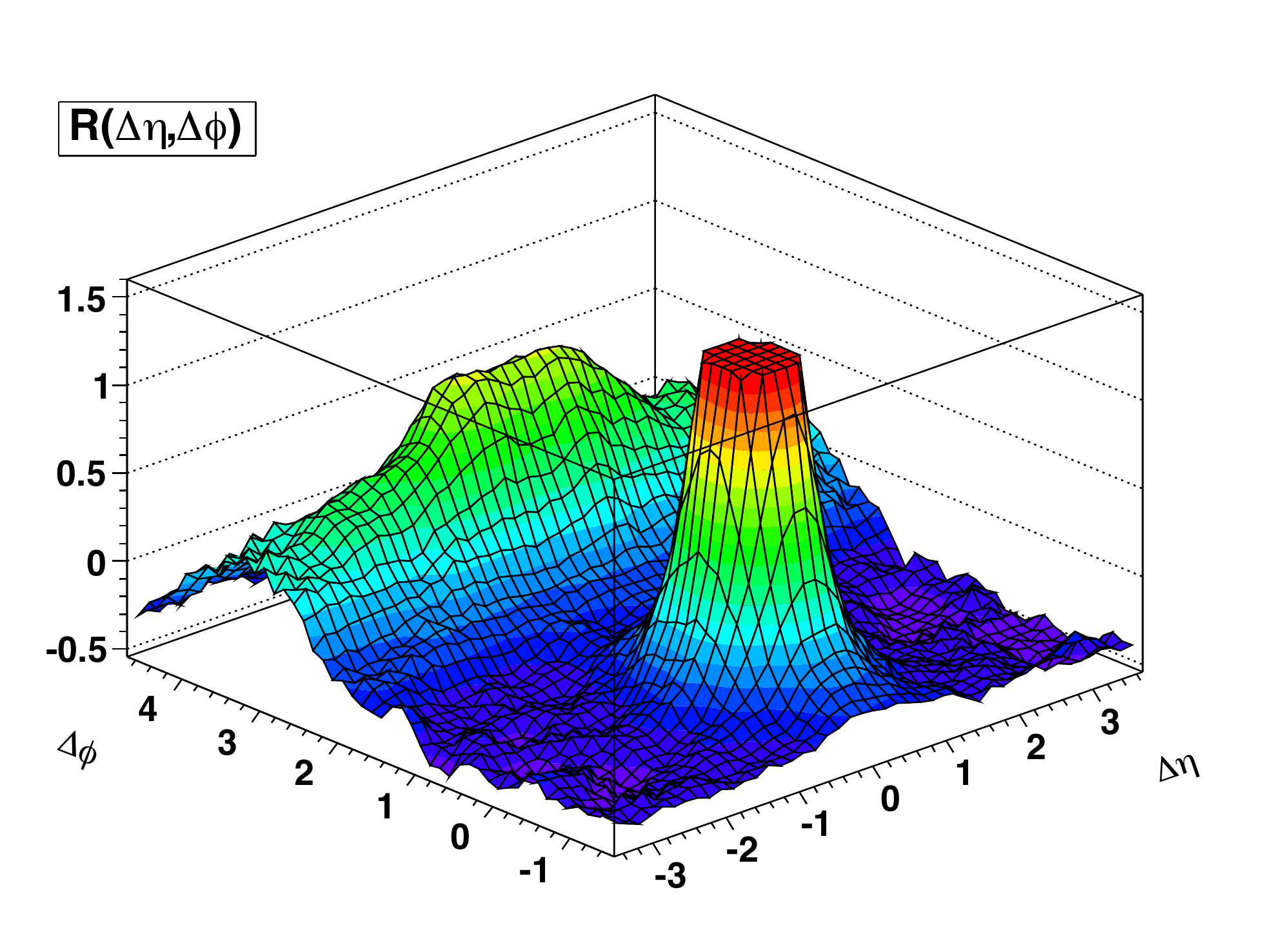}}
    \subfloat[]{\includegraphics[width=0.45\linewidth]{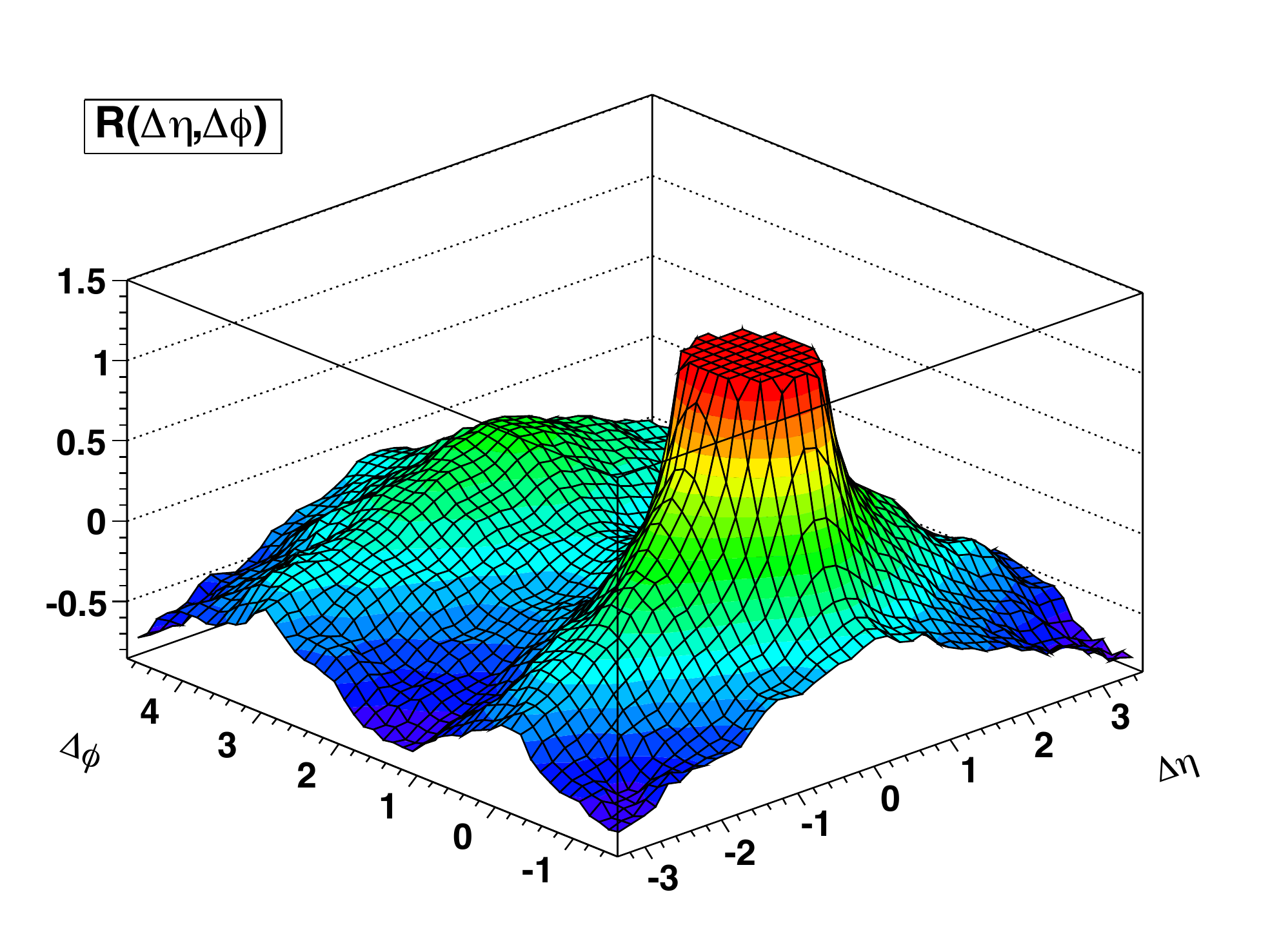}}
    \caption{2-D dihadron correlation functions from the EPOS model for high multiplicity
    events in pp collisions at \roots\ = 7 TeV (a) without hydrodynamic evolution
    and (b) with hydrodynamic evolution from Ref.~\protect\refcite{Werner:2010ss}. 
    The $p_T$ range of particles is 1--3 \GeVc.}
    \label{fig:ridge_hydro}
  \end{center}
\end{figure}

Applying the hydrodynamic approach to proton-proton collisions, it is 
feasible to assume that high multiplicity events are associated with very 
large overlap of the initial proton wavefunctions. The density profile 
of the overlapping region (mostly consisting of soft gluon fields) by 
no means has to be smooth and isotropic. Naively speaking, an oversimplified 
configuration of three ``hot spots'' initial state from
constituent quarks in a proton could generate 
an initial-state eccentricity event-by-event. Given the presence 
of final-state parton/hadron interactions (not necessarily ideal hydrodynamics 
with zero mean free path) in high multiplicity pp, the initial-state 
fluctuations would lead to a ridge-like dihadron correlation
structure, similar to that in heavy-ion collisions. The observed $p_T$ 
dependence of the ridge effect in high multiplicity pp (Fig.~\ref{fig:yieldvspt})
shows a trend of first rise and a subsequent fall at higher $p_T$. This
also resembles the behavior of collective flow phenomena in heavy ion collisions.
Fig.~\ref{fig:ridge_hydro} shows a theoretical calculation of dihadron correlations 
for high multiplicity pp using the EPOS model based on the hydrodynamic 
approach~\cite{Werner:2010ss}. The initial state consists
of multiple color flux tubes along the longitudinal direction (identical at 
different rapidity), with anisotropic energy density in the transverse 
plane. A near-side ridge structure can be clearly seen with comparable 
magnitude to the experimental data in Fig.~\ref{fig:ridge_hydro}b, 
while the effect disappears if the hydrodynamic evolution is turned off 
in the model (Fig.~\ref{fig:ridge_hydro}a). More discussions of possible elliptic and higher-order 
flow effects in pp can be found in a series of literatures in 
Refs.~\refcite{Bozek:2011if,Deng:2011at,Avsar:2011fz,Troshin:2011pd,Kisiel:2010xy,Bozek:2010pb,Avsar:2010rf}.

\begin{figure}[thb]
  \begin{center}
    \includegraphics[width=0.4\linewidth]{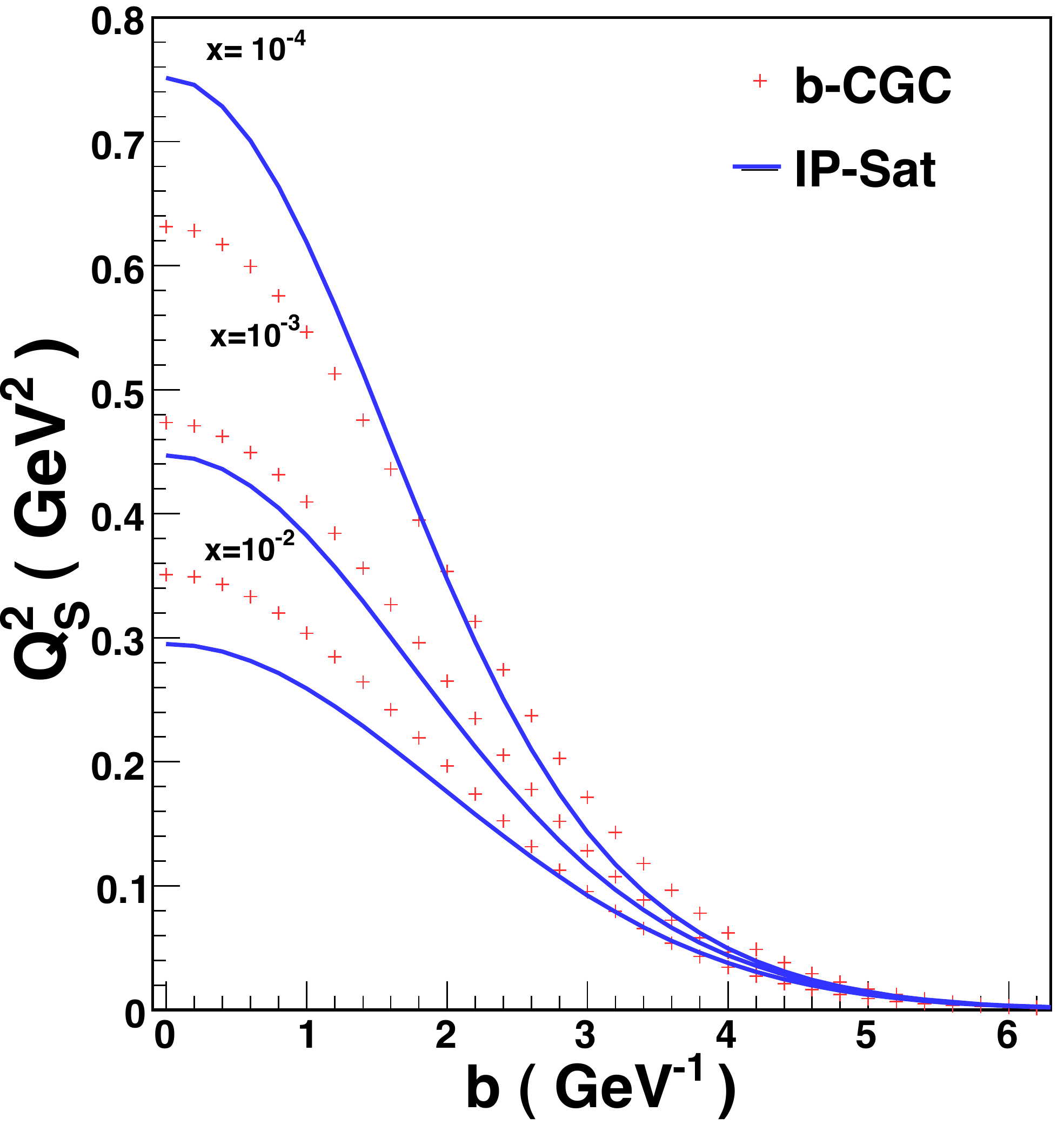}
    \hspace{2cm}
    \caption{The saturation scale as a function of impact parameter 
    in pp collisions extracted from HERA data in two different 
    saturation models~\protect\cite{Venugopalan:2010bc}.
 }
    \label{fig:saturation}
  \end{center}
\end{figure}

\begin{figure}[thb]
  \begin{center}
    \subfloat[]{\includegraphics[width=0.45\linewidth]{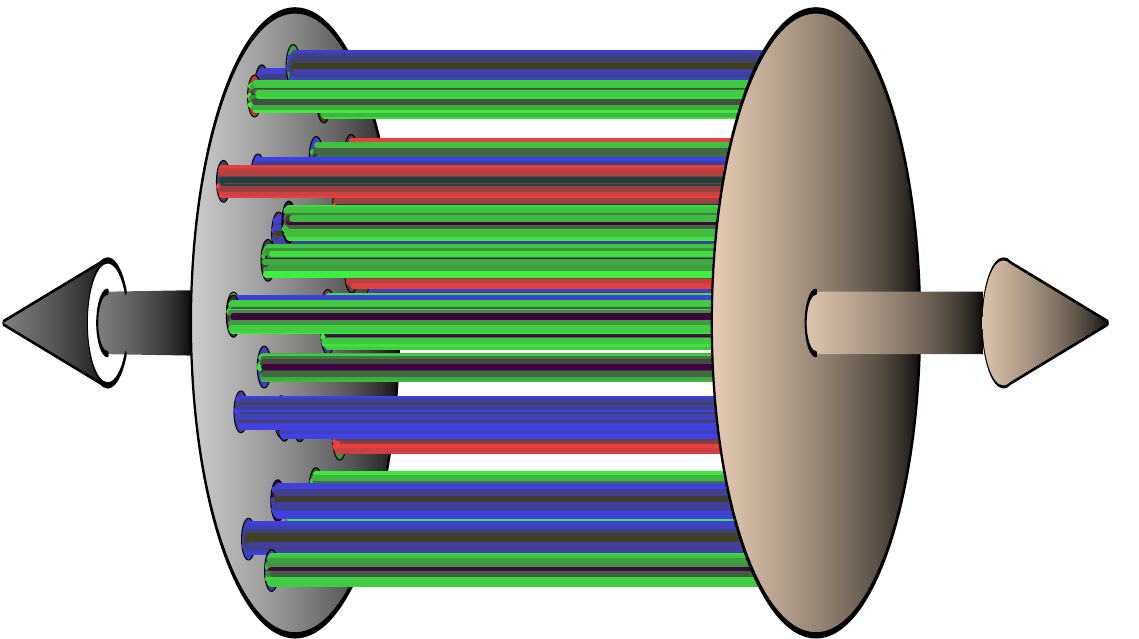}}
    \hspace{2cm}
    \subfloat[]{\includegraphics[width=0.25\linewidth]{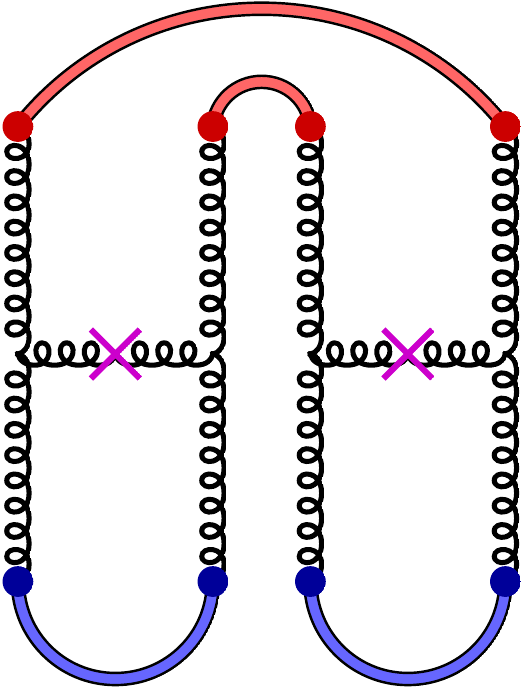}}
    \caption{(a) Glasma color flux tubes stretched between the two passing
    nuclei remnants with a typical transverse radius of $1/Q_S$; (b) Representative diagram of 
    Glasma color flux tube~\protect\cite{Dusling:2012ig}.
 }
    \label{fig:ridge_glasma}
  \end{center}
\end{figure}

\begin{figure}[thb]
  \begin{center}
    \subfloat[]{\includegraphics[width=0.50\linewidth]{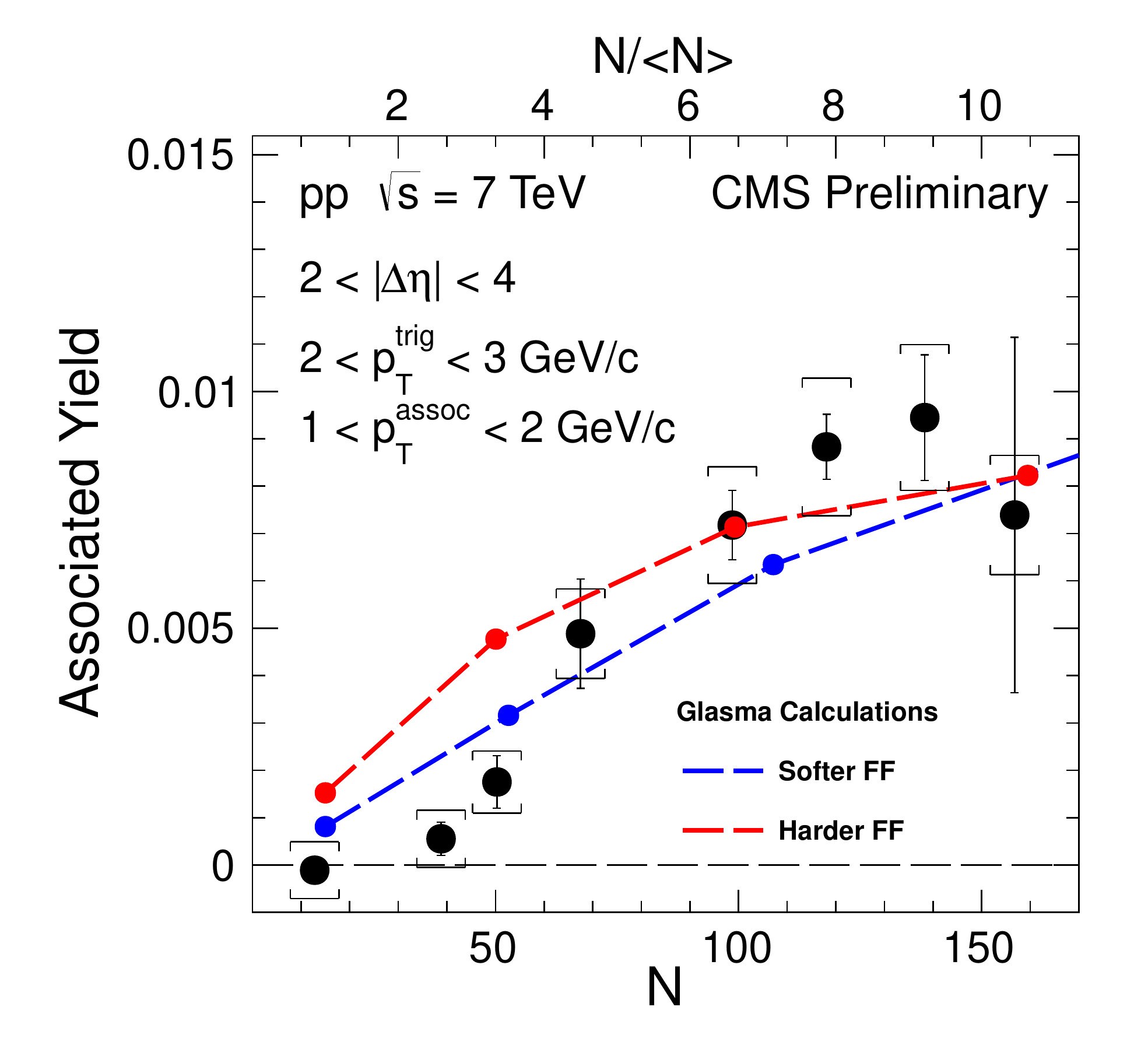}}
    \subfloat[]{\includegraphics[width=0.47\linewidth]{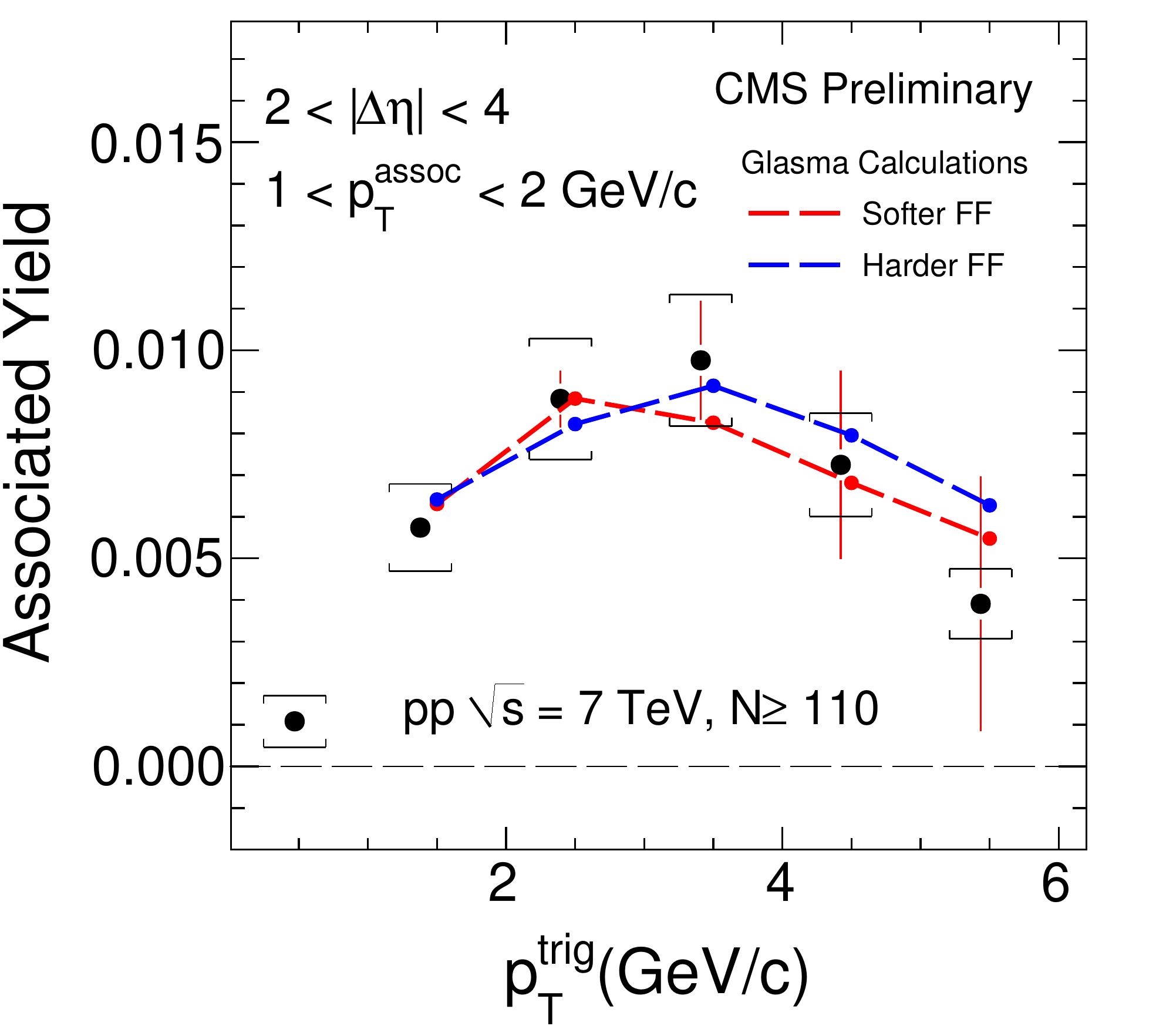}}
    \caption{
    Associated yield of the near-side ridge calculated from the Glasma model 
    (a) for four different initial saturation scales corresponding to 
    different event multiplicity and (b) as function of $p_T$
    for high multiplicity pp events~\protect\cite{Dusling:2012ig}. Different colors of the
    curves correspond to softer (blue) and harder (red) fragmentation 
    functions of the Glasma tube. Black points are data measured by the 
    CMS experiment.}
    \label{fig:ridgeyield_cgc}
  \end{center}
\end{figure}

Motivated by the concept of gluon saturation at very small $x$ in QCD, the theory of color glass 
condensate (CGC) describes the initial state of nuclear matter at 
very high energy~\cite{Iancu:2003xm,Gelis:2010nm}.
It employs the first principle approach of QCD and has been proved
to be very successful in describing a lot of observables
in high-energy nucleus-nucleus collisions. In this framework, glasma color flux 
tubes are first formed, getting stretched along the longitudinal direction as the
two nuclei are passing by, as illustrated in Fig.~\ref{fig:ridge_glasma}a.
Consequently, particles produced by the glasma tubes
are naturally correlated over long-range in rapidity. This aspect of the model 
is similar to the hydrodynamic approach presented earlier. A glasma tube has 
a typical size of $1/Q_S$ in the transverse direction, where $Q_S$ is 
the saturation scale that grows with collision energy and centrality. 
The high multiplicity events in pp are expected to sample collisions that 
correspond to very small impact parameter (or ``central'' events). 
It was argued that the saturation scale in pp increases toward smaller impact 
parameter, as shown in Fig.~\ref{fig:saturation} extracted from HERA data~\cite{Venugopalan:2010bc}.
Once $Q_S \gg \Lambda_{QCD}$ (which is the case for high multiplicity pp events), 
the perturbative approach becomes valid to describe the dynamics of 
the glasma tubes. An example diagram of ``Glasma graphs'' is shown in 
Fig.~\ref{fig:ridge_glasma}b. 
%The strings at the both ends attached to 
%the gluon propagators represent the produced hadrons~\cite{Dusling:2012ig}. Different colors
%denote strings produced from two different glasma tubes. 
Unlike the ridge
from hydrodynamics produced by the radial flow boost, an intrinsic $\Delta\phi$ collimation 
of final-state hadrons is suggested by the calculations of ``Glasma graphs'' over 
long-range in rapidity~\cite{Dumitru:2010iy,Dusling:2012ig}. The magnitude
of the intrinsic ridge from the Glasma model is found to be enough to describe 
the experimental data. No radial flow boost is needed, in contrast to
that in heavy-ion collisions. The $p_T$ and event multiplicity dependence 
of the near-side ridge yield from the Glasma model in pp show an intriguing 
agreement with the CMS data, as shown in Fig.~\ref{fig:ridgeyield_cgc}. The maximum strength of the ridge 
magnitude in particle $p_T$ occurs approximately at the saturation scale, $Q_S$.

A third category of theoretical interpretations lies along the line of
jet-induced ridge. The representative mechanisms include the energy
loss of semihard partons inducing fluctuation of local soft parton
density along their passage~\cite{Hwa:2010fj}; momentum kick
model of QCD strings pushed by the outgoing high $E_T$ partons~\cite{Wong:2011qr}.
These scenarios would predict that the ridge persists at very high $p_T$, 
which does not seem to be supported by the present data, although the statistical
uncertainties are rather limited above $p_T \sim 6$~\GeVc. Also,
as shown in Fig.~\ref{fig:highmult_2D}b, the correlation structure at $\Delta\phi \sim 0$
always consists of a narrow peak in both $\Delta\eta$ and $\Delta\phi$ from jet fragmentation
sitting on top of the long-range ridge in $\Delta\eta$. In CGC or hydrodynamic
scenarios, jet and ridge are induced by different processes, which have
no correlations in $\phi$ but only brought to have the same $\Delta\phi$
by construction. On the other hand, the jet-induced ridge would always 
be collimated with the jet in $\phi$ direction of each event. Future
studies of multiparticle correlations will be able to differentiate among
these scenarios. The scenario of jet-induced ridge
is more close to the other striking phenomena observed in heavy ion collisions,
namely the ``jet quenching''. In particular, a ridge correlation structure
is also observed recently for trigger particle $p_T$ as high as 20\GeVc\ in PbPb
collisions at the LHC~\cite{CMS_highptridge_PbPb}, where particle production must be associated 
with jet fragmentation. The common interpretation is the path length or 
initial-state geometry dependence of jet quenching effect, which generates an azimuthal
anisotropic distribution also for high $E_T$ jets but has nothing to do
with hydrodynamic flow~\cite{Peigne:2008wu,Wicks:2005gt,Jia:2010ee,Jia:2011pi,Renk:2010mf,Betz:2011tu,Betz:2012qq}. 
If a ridge signal can be observed in pp at very high $p_T$, 
it may provide intriguing evidence of jet-medium interaction,
or jet quenching, in pp collisions. Future experimental results will 
provide us the answers.
  
\section{Summary and Outlook of future directions}
\label{sec:summary}

Looking into the future, the observation of long-range near-side 
ridge correlation in high multiplicity pp collisions opens up the opportunities of 
studying very high density QCD physics in a tiny system size. Although it is 
still too early to draw any definitive conclusion on the physical original 
of the observation, a non-exhaustive list of possible further studies can 
be proposed and carried out experimentally in future program at the LHC. 
Particularly, in terms of its connection to the heavy ion physics, 
understanding the ridge phenomena in high multiplicity pp would have 
direct impact on the field of relativistic heavy ion physics, and thus 
should be extensively explored.

In order to address this question, the general approach is to investigate 
a variety of key and unique heavy-ion observables in high multiplicity 
pp events and convey a comprehensive comparison between the two systems. 
While the observation of the ridge may be regarded as a first hint of 
collective effect in pp, its properties can be examined further 
via a series of topics: 

\begin{itemize}
\item Identified hadron spectra and correlations over large rapidity range. 
If the flow effect is indeed present, a modification to the $p_T$
spectra, particularly at low $p_T$, should be observed with a dependence 
on the particle species. Meanwhile, the magnitude of the flow-induced ridge 
with identified particles would show a mass ordering as was observed in 
heavy ion collisions. This will be a critical test of the hydrodynamic scenario.

\item Studies of azimuthal correlations among multiple particles separated 
widely in rapidity would help eliminate the short-range non-flow correlations 
primarily from the jets~\cite{PrivComm_Alfred} and extract purer signature of
collective effect. Direct extraction of flow Fourier harmonics may be realized.

\item Measurements of Bose-Einstein Correlations (BEC)~\cite{Aamodt:2011kd} and heavy flavor 
production~\cite{Abelev:2012rz} in high multiplicity pp from the ALICE experiment have 
shown very interesting behaviors. The properties of these observables 
can be explored with better precision in the future and compared with 
theoretical predictions~\cite{Werner:2011fd,Wong:2011nd,Vogel:2011zz,Liu:2011dk}.
\end{itemize}
    
Furthermore, to prove the presence of jet quenching will be another crucial 
milestone forward to demonstrate the existence of the medium effect in 
high multiplicity pp collisions. The challenge lies in the difficulty of 
identifying an unambiguous reference since the hard probes themselves are 
also enhanced by requiring high multiplicity in the event. A powerful tool 
to use could be the rare electroweak probes like the isolated photon 
to calibrate the initial jet energy and study the energy distribution 
on the away side as a function of event multiplicity. As already mentioned 
in Sec.~\ref{sec:theory}, inspired by the recent observation of near-side 
ridge for very high $p_T$ particles in 2.76 TeV PbPb collisions by the 
CMS experiment at the LHC~\cite{CMS_highptridge_PbPb}, very high-$p_T$ ridge in high multiplicity
pp is probably the most promising, cleanest way to demonstrate the presence
of jet-medium interactions in these high-density pp events.

In the high-energy collisions of protons at the LHC, we are exploring an 
unprecedented territory of QCD physics under extreme condition 
in a small collision system. Although QCD has been probed with high accuracy 
in the high $p_T$ regime, its non-perturbative behavior at low $p_T$ is still poorly 
understood. It is crucial to clarify the dynamics of the high multiplicity pp
interactions and provide insight on studying the internal structure of the 
protons at much finer time and spatial scales than ever achieved before. 
Exciting new opportunities of discovering more surprises are awaiting 
us in the near future! 

\section*{Acknowledgments}

I would like to thank Dragos Velicanu for careful reading the manuscript
and valuable suggestions.

%\section*{References}

%References are to be listed in the order cited in the text in Arabic
%numerals.  They can be typed in superscripts after punctuation marks,
%e.g.~``$\ldots$ in the statement.\cite{Toimela}'' or used directly,
%e.g.~``see Ref.~\refcite{Bohr} for examples.''  Please list using the
%style shown in the following examples.  For journal names, use the
%standard abbreviations. Typeset references in 9 pt Times Roman. 
%Each reference number should consist of one reference only.

\end{document}